\documentclass[aip,reprint]{revtex4-2}

\usepackage{fdsymbol}
\usepackage{setspace}
\usepackage[utf8]{inputenc}
\usepackage[T1]{fontenc} 
\usepackage{xcolor,color,soul} 
\usepackage{graphicx}
\usepackage{bm}
\usepackage{color}
\usepackage{enumitem}
\usepackage[utf8]{inputenc}
\usepackage{wrapfig}
\usepackage{natbib}
\usepackage[hidelinks, breaklinks=true, colorlinks=true, citecolor=blue, urlcolor=blue, linkcolor=blue]{hyperref}
\usepackage{framed}
\colorlet{shadecolor}{yellow}

\draft 

\begin{document}
\title{Cross-Interaction Softness as a Route to Microphase Separation in Binary Colloidal Systems}
\author{Umesh Dhumal}
\affiliation{Department of Chemical Engineering, Indian Institute of Technology Bombay, Mumbai, Maharashtra 400076, India}
\email{umesh.dhuma007@gmail.com}

\date{\today}
	
\begin{abstract}
{ Understanding how interparticle interactions govern phase behavior is central to controlling self-organization in multicomponent soft-matter systems. In particular, the role of cross interactions between unlike components remains insufficiently understood. Here, we systematically investigate how cross-interaction character controls phase behavior in binary mixtures of hard and soft particles using coarse-grained modeling, Reference Interaction Site Model (RISM) theory, and molecular dynamics simulations. Four representative systems are examined that differ only in whether interactions between unlike particles are bounded or hard-sphere. We show that penetrable (bounded) cross interactions are both necessary and sufficient to induce microphase separation, even in the absence of attractive forces. Such systems exhibit dispersed states, macrophase separation, and microphase-separated morphologies characterized by finite-wavelength compositional ordering. In contrast, purely hard-sphere cross interactions suppress microphase separation entirely, despite strong local clustering. Comparison between theory and simulations reveals qualitative agreement in phase topology, while simulations additionally capture hierarchical and multiscale ordering near crossover regimes. These findings establish cross-interaction softness as a fundamental design principle for controlling phase behavior in multicomponent colloidal and soft-matter systems.} 
\end{abstract}
	
	
\maketitle 
	
\section*{Introduction}
\label{sec:intro}
Colloidal suspensions have attracted sustained attention due to their fundamental importance and wide-ranging applications in systems such as paints~\cite{Tadros_2011, Tadros_2013, Nappert_Reed_1991, Keddie_1997}, foods~\cite{Dickinson_1992, McClements_2015}, and pharmaceuticals~\cite{Tadros_1992, MullerPeters_1998, Mederos_2004, Tadros_2015} . Owing to their ubiquity and tunable interactions, colloids serve as versatile model systems for studying condensed matter phenomena~\cite{Likos_2001_E}. In contrast to atomic or molecular systems, the larger size of colloidal particles enables direct experimental observation of their structure and dynamics using optical and scattering techniques \cite{Gasser_2009}. This experimental accessibility allows colloidal suspensions to serve as analogues for exploring processes that are otherwise difficult to probe at atomic scales \cite{Gasser_2009}. 

Among colloidal systems, hard-sphere suspensions have been particularly instrumental in elucidating fundamental phase behaviors \cite{Sciortino_Tartaglia_2005}. Their equilibrium and nonequilibrium properties are primarily governed by the particle packing fraction, leading to phenomena such as crystallization, freezing, and glass formation \cite{Pusey_1985,Sciortino_Tartaglia_2005, Ackerson_1992, Degiorgio_1995}. Sterically stabilized poly(methyl methacrylate) (PMMA) particles, which closely approximate ideal hard spheres, have been widely employed to probe these transitions \cite{Pusey_1985, Pusey_1995}. With increasing concentration, PMMA suspensions evolve from a fluid phase to a coexistence of fluid and crystal, eventually forming fully crystalline samples \cite{Pusey_1985}. At very high packing fractions, however, crystallization can become kinetically arrested, producing amorphous, glass-like states that persist over extended timescales \cite{Pusey_1985}.

Beyond hard-sphere behavior, colloidal systems governed by effective or ultrasoft interactions—represented by bounded pair potentials that permit particle overlap—display a host of intriguing equilibrium behaviors, including clustering, crystallization, and re-entrant melting \cite{vlassopoulos2007, Overduin_2008, Likos_2001,Likos_2001_E,Curk_2012, Zhu_Lu_2010, Kalyuzhnyi_2024}. Re-entrant melting, in which a solid reverts to a fluid state upon increasing density or lowering temperature, is a characteristic feature of such ultrasoft systems \cite{Lang_1999, Travis_2024}. These phenomena are consistent with the one-component Gaussian core model (GCM) and have been observed in experimental systems such as charged colloids in nonpolar solvents and electrostatically repelling colloids, where factors such as salt concentration and pH modulate the effective interaction potential \cite{Kanai_2015, Everts_2015}. These findings underscore the central role of tunable interparticle interactions in dictating the equilibrium and nonequilibrium phase behavior of colloidal suspensions.

Binary mixtures introduce an additional level of complexity, as the interplay between softness, composition, and cross interactions can lead to qualitatively new forms of self-organization. Binary mixtures interacting via Gaussian-like potentials often avoid macroscopic demixing and instead exhibit microphase separation, where periodic density modulations emerge to minimize unfavorable contacts between dissimilar components \cite{Likos_2005}. This phenomenon parallels the behavior of block copolymers, metallic alloys, and other modulated materials \cite{BatesFredrickson_1999, Likos_2005, Carta_2011}. In particular, block copolymers have been extensively studied due to their ability to form ordered nanostructures and their broad technological relevance in drug delivery~\cite{DischerEisenberg_2002, Ahmed_2006, Discher_2007}, membrane separations~\cite{NunesPeinemann_2006, Peinemann_2007, LangKumarHillhouse_2021}, and nanoelectronics~\cite{Park_2003, Albrecht_1997, StoykovichNealey_2006}. Theoretical frameworks, beginning with Leibler’s seminal mean-field treatment of microphase separation \cite{Leibler_1980}, have provided key insights into the thermodynamics and kinetics of pattern formation in such systems.

Recent studies have further highlighted that \emph{cross interactions}—the interactions between dissimilar components—can dramatically alter phase stability and kinetics in multicomponent systems~\cite{Fortini_2016PRL, Zhou_2017PRL, SchulzKeddie_2018SoftMatter}. Depending on their nature, cross interactions can induce stratification during film drying, re-entrant transitions, and complex microstructural ordering in colloidal mixtures and Coulombic systems. Despite extensive work on bounded and ultrasoft interactions, however, a systematic understanding of how the \emph{character of cross interactions alone} governs phase behavior in binary mixtures remains incomplete. In particular, it is not yet clear whether microphase separation requires softness in the self-interactions, the cross interactions, or both. Addressing this question is essential for developing design principles for self-assembly in multicomponent colloidal and soft-matter systems.

In this work, we systematically investigate the effect of interparticle cross interactions on the phase behavior of binary mixtures. Four representative systems are considered, characterized by distinct combinations of hard and soft pair potentials. By analyzing their equilibrium phase diagrams and structural transitions using integral-equation theory and molecular dynamics simulations, we aim to elucidate how the nature and strength of cross interactions control phase stability, macrophase separation, and microphase formation. These insights contribute to a deeper understanding of interaction-driven organization in colloidal mixtures, with implications for materials design, soft condensed matter theory, and industrial formulation processes.

\section*{Modeling Approach and Methodology}
\label{sec:methods}
To characterize and design polymer-based soft matter systems, it is essential to develop a multiscale framework that bridges molecular-level interactions with mesoscopic and macroscopic behavior. Such understanding must capture polymer conformations, self-assembly, and phase separation phenomena across relevant time and length scales. At coarse-grained and mesoscopic levels, polymer molecules are typically represented as collections of interaction sites (``beads''), enabling simulations over extended spatial and temporal regimes compared with atomistic models.
	
\begin{figure}[!htb]
		\centering
		\includegraphics[width=0.9\linewidth]{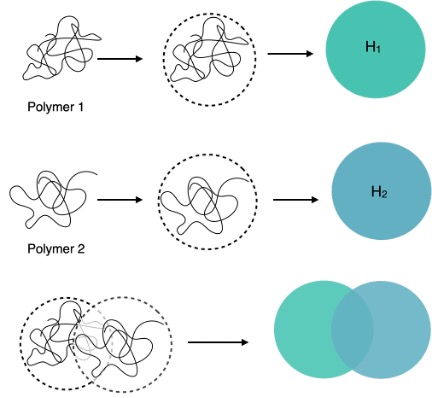}
		\caption{Schematic representation of the coarse-graining approach for a binary polymer mixture, where each polymer is modeled as a single spherical particle.}
		\label{fig:fig1}
	\end{figure}

In the present study, a highly coarse-grained description of a binary polymer mixture is employed, where each polymer chain is modeled as a single spherical particle. Interactions between polymers are represented through the overlap of these coarse-grained particles, effectively capturing entanglement and excluded-volume effects, as shown in Figure~\ref{fig:fig1}. This simplified representation facilitates systematic investigation of how interparticle interactions influence the phase behavior of binary mixtures.
	The effective interaction between particles is defined as either \emph{soft} (bounded) or \emph{hard} (impenetrable). For soft interactions, the potential is modeled using the generalized exponential model of index 3 (GEM-3):
	\begin{equation}
		U_{ij}(r) = \varepsilon_{ij} \exp\left[-\left(\frac{r_{ij}}{\sigma_{ij}}\right)^{3}\right]
		\label{eq:Uij_potential}
	\end{equation}
	Here, $\varepsilon_{ij}$ denotes the interaction strength, $\sigma_{ij}$ is the particle diameter, and $r_{ij}$ represents the interparticle separation. For all systems considered, we assume $\sigma_{11} = \sigma_{12} = \sigma_{22} = \sigma$ and $\varepsilon_{ij} = \varepsilon$ $(i,j = 1,2)$. The Fourier transform of the GEM-3 potential oscillates around zero, classifying it as a $Q^{\pm}$ potential. Systems dominated by $Q^{\pm}$ interactions typically exhibit freezing at all temperatures and form cluster crystal phases at high densities. Bounded repulsive potentials were classified according to the $Q^{+}$ and $Q^{\pm}$ criteria introduced by Likos and coworkers~\cite{Likos_2001_E, Lang_Likos_2000}. Consequently, we anticipate the emergence of cluster ordering among particles interacting via the GEM-3 potential. For hard interactions, particle overlap is prohibited, and the potential is defined as:
	\begin{equation}
		U_{ij}(r) =
		\begin{cases}
			\infty, & r \leq \sigma_{ij} \\[6pt]
			0, & r > \sigma_{ij}
		\end{cases}
		\label{eq:hard_sphere_potential}
	\end{equation}

	Four representative binary mixtures are investigated to elucidate the role of cross-interactions on phase behavior. These systems differ in whether the constituent particles interact via hard or soft potentials, and in the nature of the cross-interaction between unlike species. Specifically, two systems incorporate cross-interactions defined by a bounded GEM-3 potential: (i) a \emph{Hard--Soft--Soft (HSS) mixture}, consisting of hard and soft particles with soft cross-interactions, and (ii) a \emph{Hard--Hard--Soft (HHS) mixture}, composed of two hard-sphere species whose mutual interaction is governed by a soft potential. The remaining two systems employ cross-interactions described by an impenetrable hard-sphere potential: (iii) a \emph{Hard--Soft--Hard (HSH) mixture}, where hard and soft particles interact through a hard potential, and (iv) a \emph{Soft--Soft--Hard (SSH) mixture}, involving two soft particle species with hard cross-interactions.
	
\begin{figure}[!htb]
		\centering
		\includegraphics[width=\linewidth]{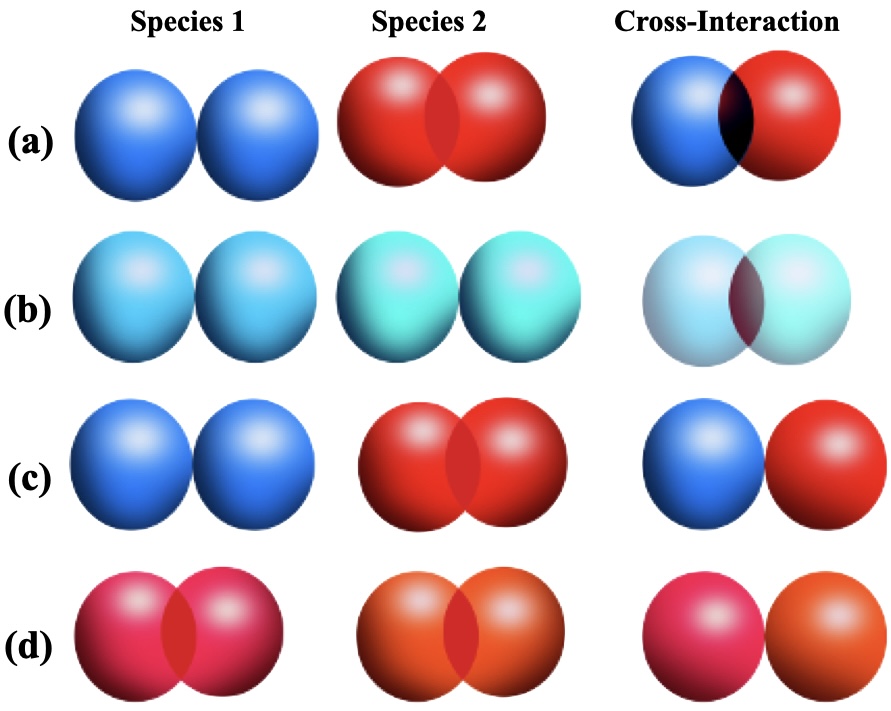}
		\caption{Schematic representation of four different mixtures of hard and soft particles. Two mixtures feature cross-interactions governed by soft potentials: (a) Hard--Soft--Soft (HSS) and (b) Hard--Hard--Soft (HHS). The other two mixtures involve cross-interactions described by hard potentials: (c) Hard--Soft--Hard (HSH) and (d) Soft--Soft--Hard (SSH).}
		\label{fig:fig2}
	\end{figure}

	A schematic representation of these four systems is shown in Figure~\ref{fig:fig2}. This classification enables a systematic comparison of how soft versus hard cross-interactions influence microphase separation, clustering, and crystallization in binary mixtures of colloidal-like particles. \par
	The structural and thermodynamic properties of the binary mixtures are analyzed using the Reference Interaction Site Model (RISM) theory~\cite{Schweizer_Curro_1997, Chandler_Andersen_1972}, which extends the Ornstein--Zernike (OZ) integral equation~\cite{Ornstein_Zernike_1914} to molecular and multicomponent systems. RISM provides a statistical-mechanical framework to describe correlations among particles by relating the total and direct correlation functions through an integral equation in reciprocal space.  
	
	For a binary mixture of spherical colloids, the RISM equation can be expressed in matrix form as:
	\begin{equation}
		\bar{H}(k) = \bar{\Omega}(k)\,\bar{C}(k)\,[\bar{\Omega}(k) + \bar{H}(k)],
		\label{eq:OZ_equation}
	\end{equation}
	where $\bar{H}(k)$, $\bar{\Omega}(k)$, and $\bar{C}(k)$ are symmetric $2\times2$ matrices representing the total, intramolecular, and direct correlation functions, respectively.  The elements of $\bar{H}(k)$ are defined as density-scaled total correlation functions,
	\begin{equation}
		H_{ij}(k) = \rho_i \rho_j h_{ij}(k),
	\end{equation}
	where $h_{ij}(k)$ is the Fourier transform of the total correlation function between particles of type $i$ and $j$, and $\rho_i$ is the number density of species $i$.  
	The intramolecular structure matrix $\bar{\Omega}(k)$ contains density-weighted intramolecular distribution functions,
	\begin{equation}
		\Omega_{ij}(k) = \sqrt{\rho_i \rho_j}\,\omega_{ij}\,\delta_{ij},
	\end{equation}
	where $\omega_{ij}$ characterizes the internal structure of a molecule (for monomeric spherical particles, $\omega_{ii}=1$ and $\omega_{ij\neq i}=0$).  
	The matrix $\bar{C}(k)$ contains direct correlation functions $C_{ij}(k)$ that encode the short-range interaction effects.  The total and direct correlation functions are linked to the real-space pair correlation function through
	\begin{equation}
		g_{ij}(r) = 1 + h_{ij}(r),
	\end{equation}
	which provides direct information about the local structural arrangement of particles in the mixture.  
	
	The Ornstein–Zernike–type equation in Eq.~\ref{eq:OZ_equation} defines a relationship between the total and direct correlation functions, yet it remains incomplete without an additional relation linking these quantities to the interparticle potential. To close this hierarchy of equations, suitable closure relations must be introduced, which approximate the effects of many-body correlations in terms of known pairwise interactions. Different closures are appropriate for distinct classes of interactions: for systems dominated by excluded-volume effects, the Percus–Yevick (PY) closure~\cite{PercusYevick1958} is typically employed, whereas for systems governed by continuous or soft potentials, the Hypernetted–Chain (HNC) closure~\cite{Ichiye1988} provides a more accurate description. Thus, we have employed the Percus--Yevick (PY) for hard interactions:
	\begin{equation}
		C_{hh}(r) = \left( 1 - e^{\beta U_{hh}(r)} \right)\left( 1 + h_{hh}(r) \right),
		\label{eq:PY_equation}
	\end{equation}
	
	For systems involving soft or mixed interactions, the Hypernetted--Chain (HNC) closure is adopted:
	\begin{equation}
		C_{ij}(r) = h_{ij}(r) - \beta U_{ij}(r) - \ln\left[ 1 + h_{ij}(r) \right],
		\label{eq:HNC_equation}
	\end{equation}
	where $\beta = 1 / (k_B T)$. The HNC closure accurately captures long-range correlations in systems with smooth or bounded potentials.  The coupled RISM and closure equations are solved self-consistently using an inexact Newton--Raphson iterative scheme implemented via the \texttt{KINSOL} algorithm. The static structure factor provides information about spatial ordering in the mixture. Once $C_{ij}(k)$ is obtained, the density-scaled static structure factor is calculated as:
	\begin{equation}
		S'(k) = \left[ I - \Omega(k) C(k) \right]^{-1} \Omega(k),
		\label{eq:SK_equation}
	\end{equation}
	and the partial structure factors are given by:
	\begin{equation}
		S_{ij}(k) = \frac{S'(k)}{\sqrt{\rho_i \rho_j}}.
	\end{equation}
	Large peaks in $S'(k)$ at non-zero wavevectors ($k > 0$) indicate \emph{microphase separation} with a characteristic periodicity $L = 2\pi/k$, while peaks at $k = 0$ signify \emph{macrophase separation}. According to the Hansen--Verlet criterion ($S_{ij}(k) \gtrsim 2.85$)~\cite{HansenVerlet1969}, such peaks correspond to the onset of freezing or structural ordering in the system. Although originally developed for simple atomic liquids, the Hansen–Verlet criterion has been widely used as a qualitative indicator of ordering and instability in soft-matter systems, including colloidal and polymeric fluids.  Overall, the RISM framework allows us to predict how the balance between hard and soft interactions, combined with cross-interaction effects, governs structural correlations and phase transitions in binary colloidal mixtures. 
	
To validate the theoretical predictions, molecular dynamics (MD) simulations were performed using the \texttt{LAMMPS} package~\cite{Plimpton_1995}. Periodic boundary conditions were applied in all three spatial directions, and the equations of motion were integrated using the Velocity--Verlet algorithm with a timestep of $\Delta t = 10^{-3}$--$10^{-4}\,\tau$, where $\tau = \sqrt{m\sigma^2/\varepsilon}$ is the characteristic time unit of the system. This choice of timestep ensures numerical stability and accurate resolution of both short-range collisions and collective structural relaxation.

All simulations were carried out in the canonical (NVT) ensemble, with temperature controlled using a Nosé--Hoover thermostat~\cite{Nose1984,Hoover1985}. The Nosé--Hoover scheme enables proper sampling of the canonical ensemble while preserving realistic dynamical correlations, making it well suited for studying structural ordering and phase behavior in soft-matter systems. The system was initially prepared in a crystalline configuration and melted at a high temperature $T = 5\,\varepsilon/k_B$ for $5\times10^5$ integration steps to eliminate memory of the initial structure and obtain a homogeneous fluid state. Subsequently, annealing was performed by gradually reducing the temperature to $T = 1\,\varepsilon/k_B$ over $2\times10^7$--$10\times10^7$ steps, allowing the system to explore low-energy configurations and develop equilibrium structural correlations. After annealing, the system was equilibrated at the target temperature for an additional $2\times10^8$--$8\times10^8$ steps to ensure full relaxation, particularly near phase-transition regimes where equilibration times can be long. Production runs of $1\times10^8$ steps were then carried out to accumulate statistically reliable data.

Radial distribution functions obtained from molecular dynamics simulations were compared directly with the corresponding predictions from RISM theory to assess the consistency between numerical and theoretical descriptions of structural correlations. Within the theoretical framework, static structure factors were used to identify macrophase and microphase separation through the presence of peaks at zero or finite wave vectors. In the simulations, radial distribution functions together with visual inspection of equilibrated configurations provided complementary real-space identification of dispersed, microphase-separated, and macrophase-separated states, which we now use to systematically analyze the four model systems introduced above.

\section*{Results and Discussion}
	
\subsection*{Bounded potential between unlike particles}
	In this section, we investigate the phase behavior of binary mixtures in which the interactions between unlike particles are governed by a bounded soft potential of the GEM-3 form. Two representative systems are considered: (i) the \emph{Hard--Soft--Soft (HSS) mixture}, composed of hard and soft particles where the cross-interaction between dissimilar species is described by the GEM-3 potential, and (ii) the \emph{Hard--Hard--Soft (HHS) mixture}, consisting of two hard-sphere components whose mutual cross-interaction is likewise defined by a soft bounded potential. These systems were selected to elucidate how the interplay between excluded-volume effects and penetrable interactions influences structural organization, microphase formation, and phase separation in multicomponent colloidal mixtures.
	
\begin{figure}[!h]
	\centering
	\includegraphics[width=\linewidth]{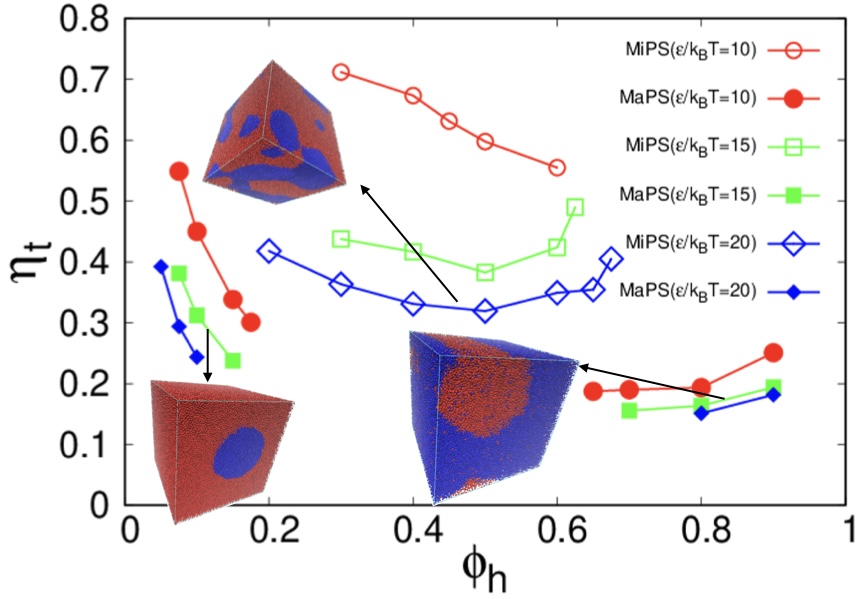}
	\caption{Spinodal phase diagram of the Hard--Soft--Soft (HSS) system showing macrophase separation at low and high fractions of hard colloids and microphase separation at intermediate hard-colloid fractions ($\phi_h$), for various total packing fractions ($\eta$) and interaction strengths ($\varepsilon/k_BT$), as predicted by RISM theory. Representative molecular dynamics snapshots illustrating the corresponding morphologies are also shown. Figure adapted from our previous study by Erigi \emph{et al.}~\cite{Erigi_2023}.}
	\label{fig:fig3}
\end{figure}
	
	\subsubsection*{Phase behavior of the Hard--Soft--Soft (HSS) system}
	
	The phase behavior of the Hard--Soft--Soft (HSS) system has been investigated in detail in our previous work~\cite{Erigi_2023} using PRISM/RISM theory in combination with molecular dynamics simulations. In this system, hard colloids interact with soft coarse-grained macromolecules through a bounded repulsive potential, giving rise to a rich interplay between excluded-volume effects and penetrable interactions. As summarized in Fig.~\ref{fig:fig3}, the system exhibits distinct phase regimes depending on composition, density, and interaction strength.
	
	At low and high fractions of hard colloids, the system undergoes \emph{macrophase separation}, characterized by the formation of extended hard-rich and soft-rich regions. In these regimes, the asymmetry in component concentration and effective interactions favors bulk demixing as the dominant route to free-energy minimization. In contrast, at intermediate hard-colloid fractions, the system displays pronounced \emph{microphase separation}, where complete demixing is suppressed and the mixture instead organizes into compositionally modulated domains with bicontinuous or clustered morphologies. These microstructured states are stabilized over a wide range of packing fractions and interaction strengths.
	
	\begin{figure*}[t]
		\centering
		\includegraphics[width=0.6\textwidth,angle=270]{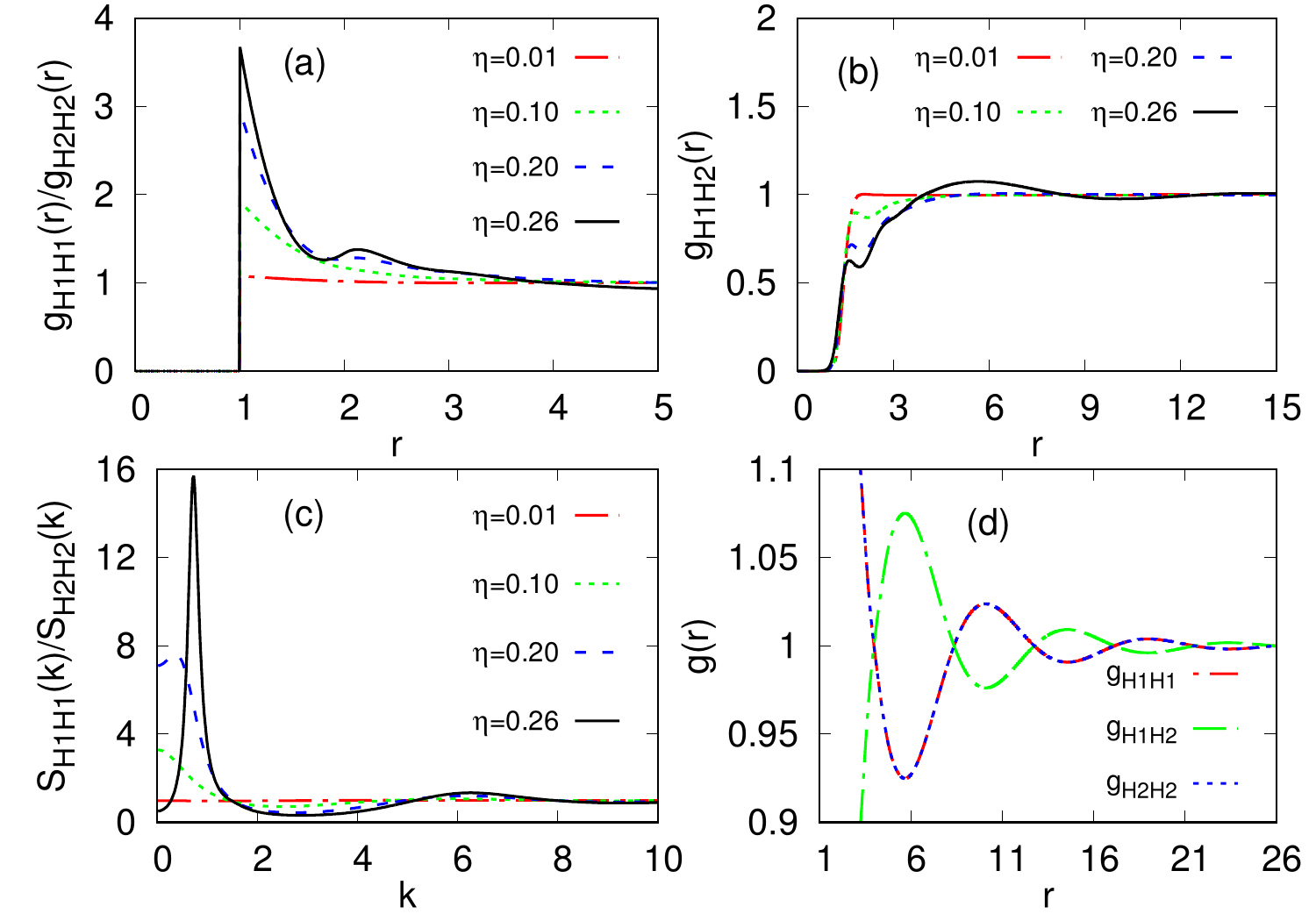}
		\caption{(a) Radial distribution functions $g(r)$ for $H_1$--$H_1$ and $H_2$--$H_2$ pairs, (b) radial distribution functions for unlike pairs ($H_1$--$H_2$), (c) structure factors $S_{H_1H_1}(k)$ and $S_{H_2H_2}(k)$ showing the emergence of a finite-$k$ peak with increasing $\eta$, and (d) combined real-space pair correlations at $\eta=0.26$. Long-range oscillations indicate compositional ordering and domain periodicity.}
		\label{fig:fig4}
	\end{figure*}

	The emergence of microphase separation in the HSS system can be directly attributed to the bounded nature of the GEM-3 interaction, which permits partial overlap between hard and soft components. This penetrability penalizes sharp interfaces and disfavors complete phase separation, while simultaneously promoting finite-wavelength compositional ordering. Such behavior is consistent with the general framework for soft and ultrasoft matter systems, where competition between hard-core exclusion and bounded repulsion leads to modulated phases rather than macroscopic demixing~\cite{Likos_2001_E,Overduin_2008}. Importantly, the HSS system demonstrates that microphase separation can arise in the absence of attractive interactions, driven purely by repulsive interactions with distinct length scales.
	
	The phase behavior of the HSS system therefore provides a crucial reference for the present study. In particular, it motivates the central question addressed here: whether similar microphase-separated states can be stabilized when \emph{both} components interact as hard particles, and bounded interactions are introduced only through the cross-interaction. This consideration naturally leads to the investigation of the Hard--Hard--Soft (HHS) system, where symmetry between the components is restored at the level of self-interactions, and the role of cross-interaction softness can be isolated and examined systematically.

	In the present work, we extend this analysis by combining RISM theory with molecular dynamics (MD) simulations to explore the detailed structural correlations, domain formation, and phase transitions in remaining 3 systems defined earlier: \emph{HHS, SHH, SSH}.  The occurrence of microphase and macrophase separation is identified through both the static structure factor and radial distribution functions, and further confirmed through MD simulations. The combined theoretical and simulation results provide a consistent picture of how cross-interactions control domain morphology and structural ordering in binary mixtures of hard and/or soft particles.
	
\subsubsection*{Phase behavior of the Hard--Hard--Soft (HHS) system}
	
	In the Hard--Hard--Soft (HHS) system, particles $H_1$ and $H_2$ interact with identical species via hard-sphere potentials, which strictly prohibit overlap between like particles. In contrast, the cross-interaction between unlike particles ($H_1$--$H_2$) is described by a bounded GEM-3 potential, allowing partial overlap between dissimilar species. This asymmetry between self- and cross-interactions introduces a competition between excluded-volume repulsion and penetrable overlap, a mechanism that has been shown to generate finite-wavelength instabilities and microphase-separated states in systems governed by bounded or ultrasoft interactions~\cite{Likos_2001_E,Overduin_2008}. The resulting structural and phase behavior is examined as a function of the $H_2$ fraction ($\phi_{H_2}$), total packing fraction ($\eta$), and interaction strength ($\varepsilon/k_BT$) using RISM theory, with validation from molecular dynamics (MD) simulations.
	
	At low packing fractions, the mixture remains structurally homogeneous, with weak correlations in all pair distribution functions (Fig.~\ref{fig:fig4}a). As $\eta$ increases, excluded-volume constraints promote short-range ordering among like particles, reflected in the growth of the first-neighbour peaks in $g_{H_1H_1}(r)$ and $g_{H_2H_2}(r)$. Importantly, the cross-correlation function $g_{H_1H_2}(r)$ develops pronounced oscillations extending over several particle diameters at higher $\eta$ (Fig.~\ref{fig:fig4}b). These oscillations signal the emergence of alternating $H_1$-rich and $H_2$-rich domains, indicating the onset of compositional modulation rather than simple depletion or bulk demixing.
	
	The corresponding structure factors (Fig.~\ref{fig:fig4}c) provide further evidence for this transition. With increasing $\eta$, a dominant peak emerges at a finite wavevector $k^*>0$, whose height satisfies the Hansen–Verlet criterion ($S_{ij}(k^*)\gtrsim2.85$), confirming the development of an ordered state. The associated characteristic length scale, $L=2\pi/k^*$, is approximately 8–10 particle diameters, consistent with the domain sizes observed in the real-space correlations (Fig.~\ref{fig:fig4}d). Such finite-wavelength ordering is a hallmark of systems governed by $Q^{\pm}$-type bounded potentials, where the oscillatory Fourier transform of the interaction favors microphase separation over macrophase demixing~\cite{Likos_2001_E}. Similar behavior has been reported in binary mixtures of ultrasoft particles, where penetrable cross-interactions stabilize periodic density modulations~\cite{Overduin_2008}.
	
	\begin{figure}[!htb]
		\centering
		\includegraphics[width=\linewidth]{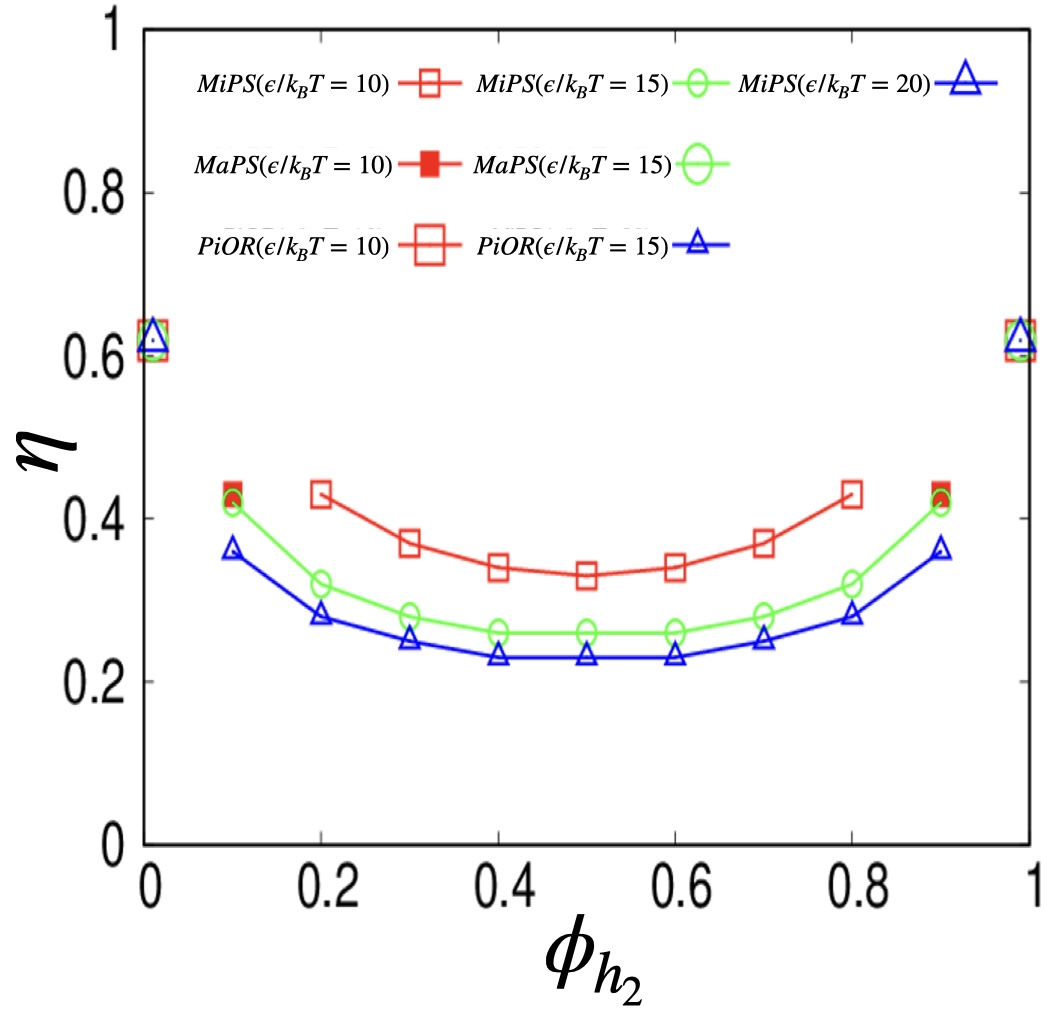}
		\caption{Phase diagram of the HHS system in the $\eta$--$\phi_{H_2}$ plane obtained from RISM theory. Macrophase separation (MaPS) occurs at extreme compositions for weak cross-interactions, while microphase separation (MiPS) dominates at intermediate $\phi_{H_2}$. Increasing $\varepsilon/k_BT$ stabilizes MiPS and suppresses MaPS.}
		\label{fig:fig5}
	\end{figure}
	
	The phase behavior summarized in the $\eta$--$\phi_{H_2}$ diagram (Fig.~\ref{fig:fig5}) highlights the central role of cross-interaction strength in determining the system topology. At weak interaction strength ($\varepsilon/k_BT=10$), macrophase separation is observed at very low and very high $\phi_{H_2}$, reflecting the dominance of entropic demixing when the bounded overlap between unlike particles is insufficient to stabilize mixed configurations. In this regime, the system minimizes free energy by separating into $H_1$-rich and $H_2$-rich bulk domains, analogous to demixing tendencies in asymmetric hard-sphere mixtures.
	
	As the interaction strength increases ($\varepsilon/k_BT=15$ and $20$), macrophase separation is progressively suppressed and replaced by microphase separation across a broad range of compositions. Stronger bounded cross-interactions penalize complete demixing while promoting partial overlap between unlike species, leading to periodic modulation of composition rather than macroscopic segregation. This behavior closely parallels observations in binary ultrasoft mixtures~\cite{Overduin_2008} and in mixtures of hard colloids and soft coarse-grained macromolecules~\cite{Erigi_2023}, where microphase separation emerges from the competition between hard-core exclusion and soft, penetrable interactions. The systematic expansion of the microphase-separated (MiPS) region with increasing $\varepsilon$ therefore underscores interaction softness as a key control parameter governing self-organization in otherwise hard-particle mixtures.
	
	\begin{figure}[!htb]
		\centering
		\includegraphics[width=\linewidth]{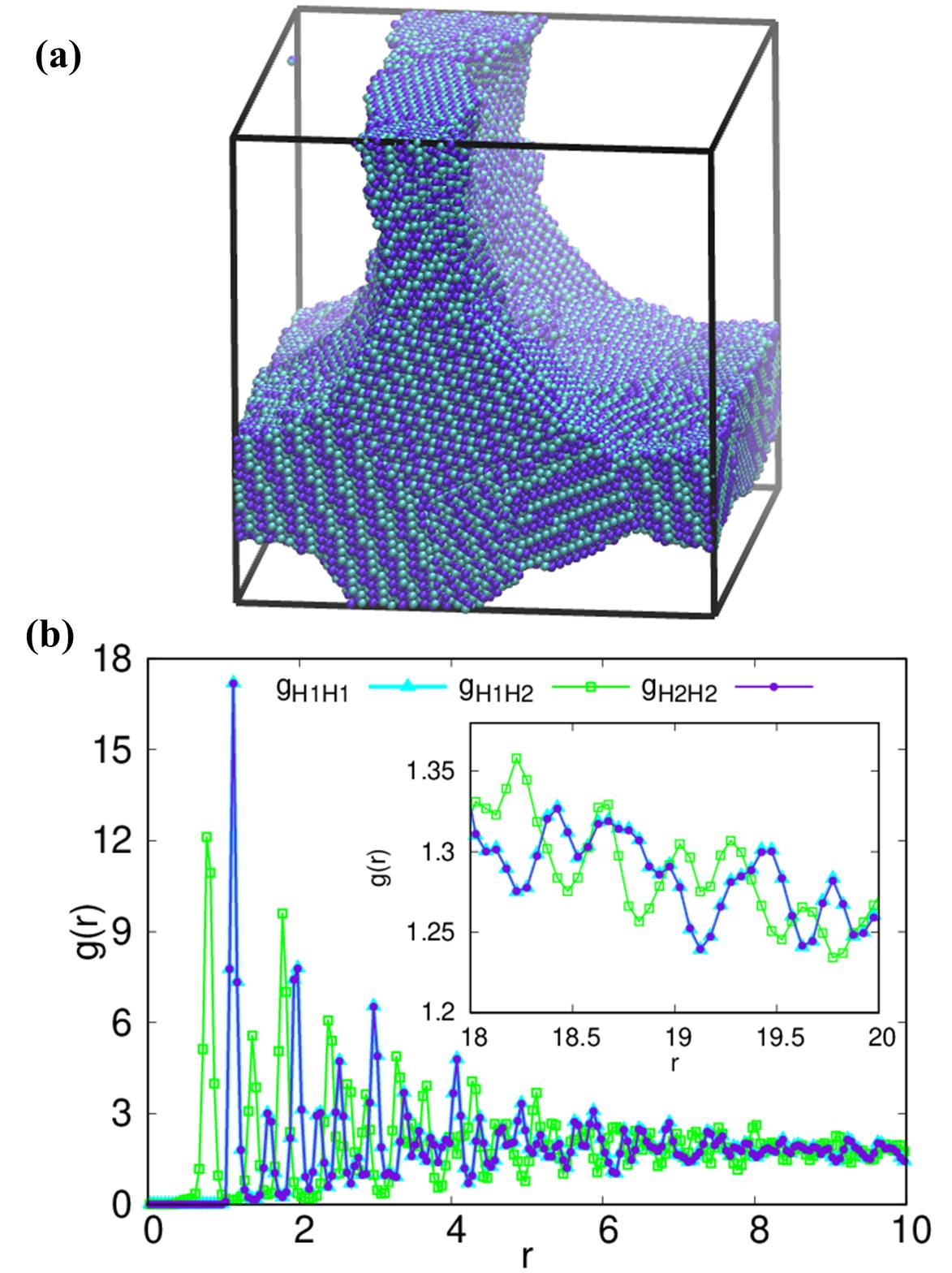}
		\caption{(a) MD snapshot for $m=3$, $\varepsilon/k_BT=15$, $\phi_{H_2}=0.5$, and $\eta=0.42$ with simulation box size 60, showing alternating $H_1$- and $H_2$-rich domains. (b) Corresponding radial distribution functions. The inset highlights short-range compositional ordering, while the slow decay of oscillations reflects long-range structural correlations.}
		\label{fig:fig6}
	\end{figure}
	
	Direct MD simulations further elucidate the nature of ordering in this regime. At $m=3$, $\varepsilon/k_BT=15$, $\phi_{H_2}=0.5$, $\eta=0.42$ and simulation box size of 60, the equilibrium configuration (Fig.~\ref{fig:fig6}a) exhibits an interconnected network of alternating $H_1$- and $H_2$-rich regions. Analysis of the radial distribution functions (Fig.~\ref{fig:fig6}b) reveals two distinct length scales. The most pronounced compositional ordering occurs over short distances of approximately one to two particle diameters, as highlighted in the inset of Fig.~\ref{fig:fig6}b. This local ordering reflects microphase separation at the particle scale, driven by the bounded cross-interaction that favors periodic local arrangements of unlike particles.
	
	At the same time, the simulation snapshot shows a clear tendency toward large-scale densification: the microstructured material collapses into a compact aggregate occupying only part of the simulation box. For a box of linear dimension $L=60$, the resulting structure spans a substantial fraction of the box (of order $L/2$), leaving an extended low-density region. This global shrinkage indicates a macrophase-separation tendency on long length scales, coexisting with local microphase ordering. The MD results therefore reveal hierarchical organization, in which short-range compositional modulation is embedded within a slowly evolving, macroscopically segregated structure.
	
	This multiscale behavior differs quantitatively from the characteristic length scale inferred from RISM theory, which predicts microphase separation with a domain spacing of roughly 8--10 particle diameters based on the finite-$k$ instability. This mismatch reflects the mean-field nature of integral-equation approaches, which accurately capture the dominant instability wavelength in a homogeneous reference state but neglect fluctuation-driven effects, kinetic pathways, and finite-size constraints that become apparent in particle-resolved simulations. In particular, MD simulations naturally access long-wavelength density fluctuations and slow structural rearrangements during annealing, which can promote large-scale aggregation beyond the length scale predicted by RISM.
	
	Importantly, the coexistence of local microphase separation and global densification does not indicate a breakdown of the theoretical description, but rather places the system near a crossover regime where competing interactions generate ordering on multiple length scales. Similar hierarchical and multiscale structures have been reported in simulations of soft-matter systems with penetrable or ultrasoft interactions~\cite{Likos_1998_PRE_PenetrableSpheres, Likos_2001_PRE_Criterion, Overduin_2008, Zhang_Charbonneau_Mladek_2010_PRL, Liu_Xi_2019_COCIS, Munao_2022_SoftMatter, Erigi_2023}.
	
	To assess the robustness of this behavior, we repeated the MD simulations using independently generated random initial configurations and identical thermodynamic conditions. In all cases, the system again exhibited local microphase ordering at short length scales together with large-scale shrinkage of the structure, although the detailed domain morphology varied. This confirms that the observed multiscale ordering is not an artifact of a particular initialization protocol. Representative configurations from these independent runs are provided in the Appendix.
	
	To summarize, the Hard--Hard--Soft system demonstrates that introducing a bounded cross-interaction between otherwise hard particles fundamentally reshapes the phase behavior of binary mixtures. While purely hard-sphere systems undergo freezing or vitrification at high densities~\cite{Pusey_1985}, penetrable cross-interactions stabilize microphase-separated states characterized by local compositional ordering. At the same time, residual hard-sphere exclusion can drive large-scale densification, leading to hierarchical structures that combine microphase separation at short length scales with macrophase tendencies at long length scales. These results establish cross-interaction softness as a decisive control parameter for self-organization in hard-particle mixtures and highlight a robust route to microphase separation without invoking attractive interactions.
	
		\begin{figure*}[t]
		\centering
		\includegraphics[width=0.6\textwidth,angle=270]{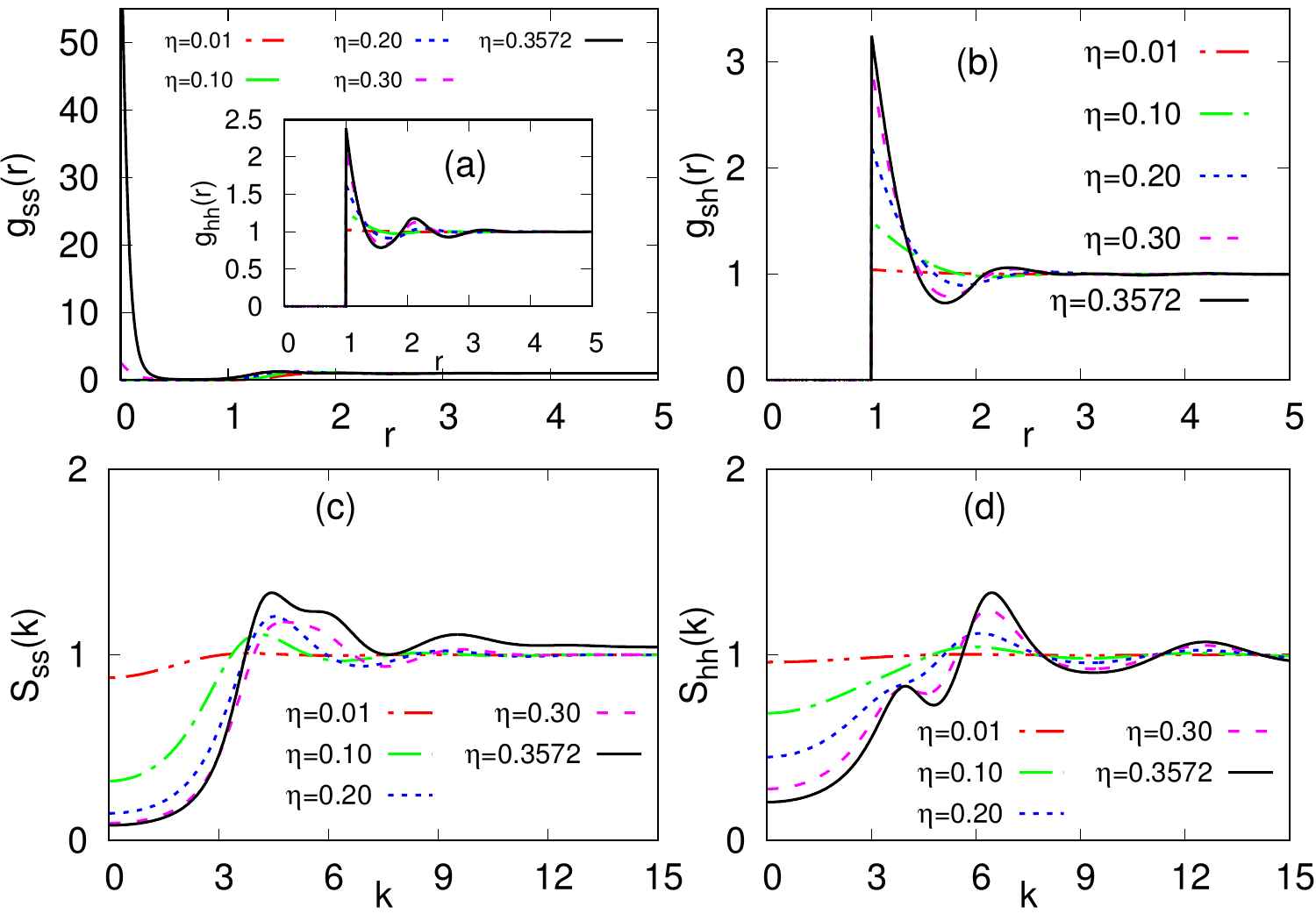}
		\caption{(a) Radial distribution functions $g(r)$ for soft--soft pairs (inset: hard--hard pairs) in the Hard--Soft--Hard system at $m=3$, $\varepsilon=15$, and $\phi_h=0.5$ for varying packing fraction $\eta$. (b) Radial distribution function for unlike pairs (soft--hard), showing excluded overlap due to hard-sphere cross interactions. (c) Partial structure factor $S_{ss}(k)$ for soft--soft pairs. (d) Partial structure factor $S_{hh}(k)$ for hard--hard pairs. The absence of a pronounced finite-$k$ peak indicates no microphase ordering across the $\eta$ range shown.}
		\label{fig:fig7}
	\end{figure*}
	\subsection*{Hard-sphere cross interactions between unlike particles}
		
In the preceding sections, we demonstrated that bounded cross-interactions between unlike particles can drive rich phase behavior, including microphase separation and hierarchical ordering, even in the absence of attractive forces. We now turn to the complementary scenario in which the cross-interactions between unlike particles are described by a hard-sphere potential. This choice allows us to isolate the role of cross-interaction softness by directly contrasting systems with identical self-interactions but qualitatively different cross-interaction character. As bounded cross interactions promote finite-wavelength instabilities and microphase separation, we now treat hard-sphere cross interactions as a controlled negative test in which overlap between unlike species is strictly forbidden.From a physical perspective, this restriction removes the possibility of entropy redistribution through partial overlap, thereby suppressing the coupling between local clustering and long-wavelength compositional fluctuations that underlies microphase separation in bounded systems.

Two representative systems are considered: (i) the \emph{Hard--Soft--Hard (HSH)} system, consisting of hard and soft particles whose cross-interaction is purely hard-sphere, and (ii) the \emph{Soft--Soft--Hard (SSH)} system, consisting of two soft particle species interacting via bounded potentials within each species but hard-sphere interactions between unlike species. These systems test whether microphase separation can arise in the absence of penetrable cross interactions.

\subsection*{Phase behavior of the Hard--Soft--Hard (HSH) system}
	
	In the Hard--Soft--Hard (HSH) system, hard particles interact with one another via hard-sphere potentials, while soft particles interact among themselves through a bounded GEM-3 potential that allows partial or complete overlap. In contrast to the Hard--Soft--Soft system discussed earlier, the cross-interaction between hard and soft particles is here described by a hard-sphere potential, preventing overlap between unlike species. This modification removes the penetrability that was previously shown to stabilize microphase-separated states and allows us to assess whether clustering within a single component is sufficient to generate long-range compositional ordering.

	The radial distribution functions in Fig.~\ref{fig:fig7} reveal distinct local structuring for each species. As shown in Fig.~\ref{fig:fig7}(a), soft--soft correlations exhibit a pronounced peak at $r=0$, reflecting extensive particle overlap and the formation of soft-particle clusters enabled by the bounded GEM-3 interaction. Such clustering is a well-known feature of ultrasoft and penetrable particle systems~\cite{Likos_2001_E,Mladek_2006_PRL_ClusterPhases}. In contrast, the hard--hard correlations (inset of Fig.~\ref{fig:fig7}(a)) display a contact peak at $r=\sigma$, characteristic of excluded-volume packing without any indication of medium- or long-range order. The soft--hard radial distribution function in Fig.~\ref{fig:fig7}(b) shows a sharp contact peak at $r=\sigma$ and vanishes for $r<\sigma$, confirming that overlap between unlike particles is completely suppressed by the hard-sphere cross-interaction.
	
	Despite the presence of pronounced clustering among the soft particles, the corresponding structure factors in Fig.~\ref{fig:fig7}(c,d) show no pronounced peaks at finite wavevectors for either species across the entire range of packing fractions studied. The absence of a finite-$k$ instability indicates that neither microphase separation nor macrophase separation occurs. Instead, the system remains in a dispersed state, with local clusters of soft particles embedded within an otherwise homogeneous mixture. This demonstrates that clustering within one component alone is insufficient to induce microphase separation when the cross-interaction forbids overlap and thus prevents coupling between local clustering and long-wavelength density modulations~\cite{Overduin_2008,Erigi_2023}.
	
	\begin{figure}[!ht]
		\centering
		\includegraphics[width=\linewidth]{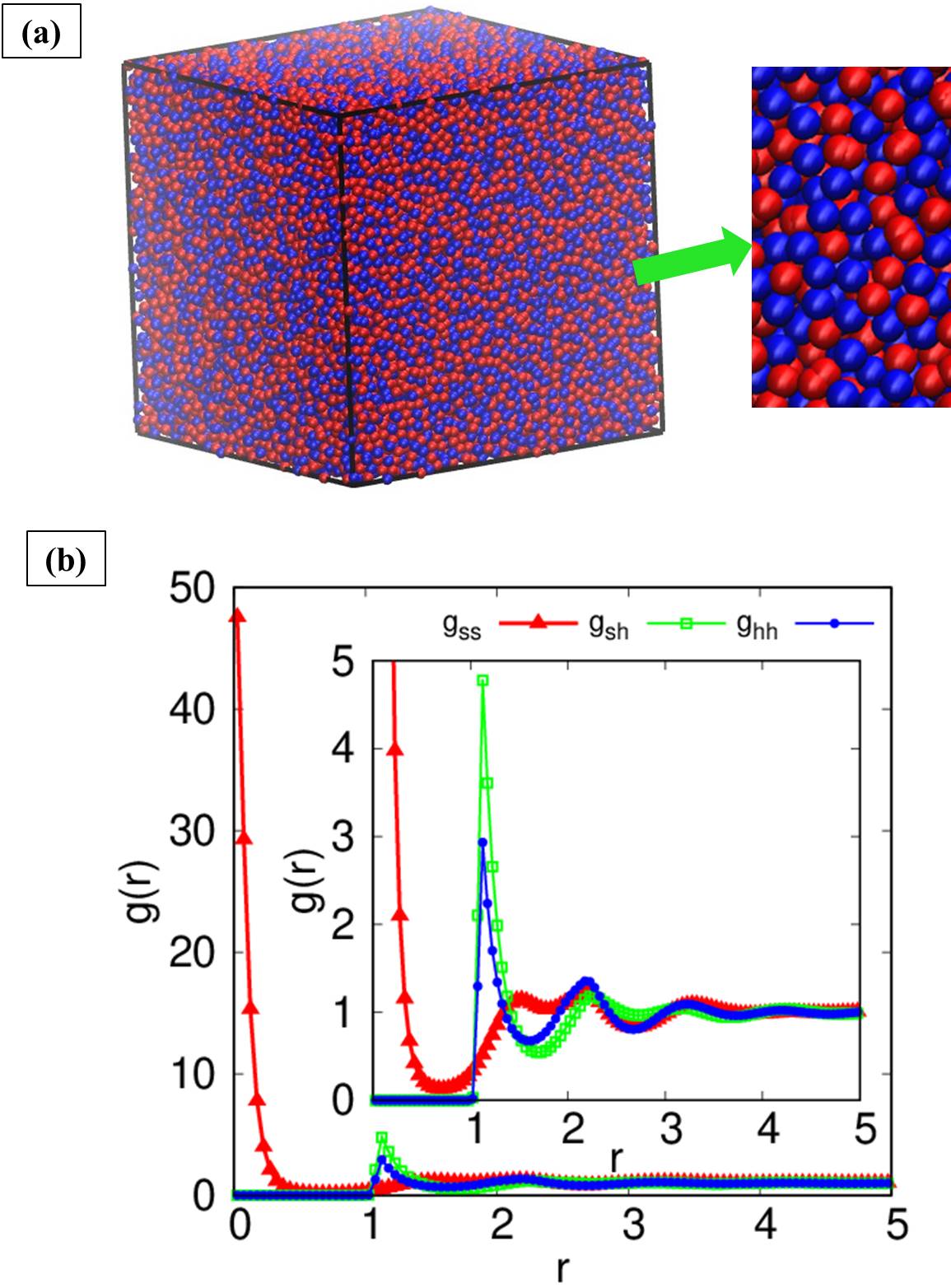}
		\caption{Molecular dynamics results for the Hard--Soft--Hard system at $m=3$, $\varepsilon = 15$, $\phi_h = 0.5$, $\eta = 0.4$ and simulation box of size 60. (a) Equilibrium snapshot showing dispersed hard particles and overlapping soft-particle clusters. (b) Corresponding radial distribution functions.}
	\end{figure}
	
	Molecular dynamics simulations shown in Fig.~8 corroborate the theoretical predictions and provide a real-space view of the equilibrium structure. The simulation snapshot (Fig.~8(a)) confirms that soft particles form overlapping clusters, while hard particles remain spatially dispersed throughout the simulation box. Importantly, these clusters do not organize into periodic or extended compositional patterns. The corresponding radial distribution functions (Fig.~8(b)) show a large peak at $r=0$ for soft--soft pairs, consistent with cluster formation, while soft--hard and hard--hard correlations exhibit contact peaks at $r=\sigma$ characteristic of hard-sphere exclusion. Beyond a few particle diameters, all correlation functions rapidly decay toward unity, indicating the absence of long-range density modulation or domain formation.
	
	The consistency between RISM theory and MD simulations underscores that hard-sphere cross interactions stabilize a dispersed phase even in the presence of strong local clustering among soft particles. These results highlight that penetrability of the \emph{cross interaction}, rather than clustering within individual components, is the decisive ingredient for the emergence of microphase separation in binary mixtures of hard and soft particles.

\subsection*{Phase behavior of the Soft--Soft--Hard (SSH) system}
	
	We next consider the Soft--Soft--Hard (SSH) system, in which two soft particle species interact via bounded GEM-3 potentials within each species, while the cross-interaction between unlike soft particles is described by a hard-sphere potential. As a result, particles of the same species are fully penetrable, whereas mutual overlap between $S_1$ and $S_2$ particles is strictly forbidden. This system provides a stringent test of whether microphase separation can arise when both components are intrinsically ultrasoft but penetrability is removed at the level of the cross interaction.
	
	\begin{figure*}[t]
		\centering
		\includegraphics[width=0.6\textwidth,angle=270]{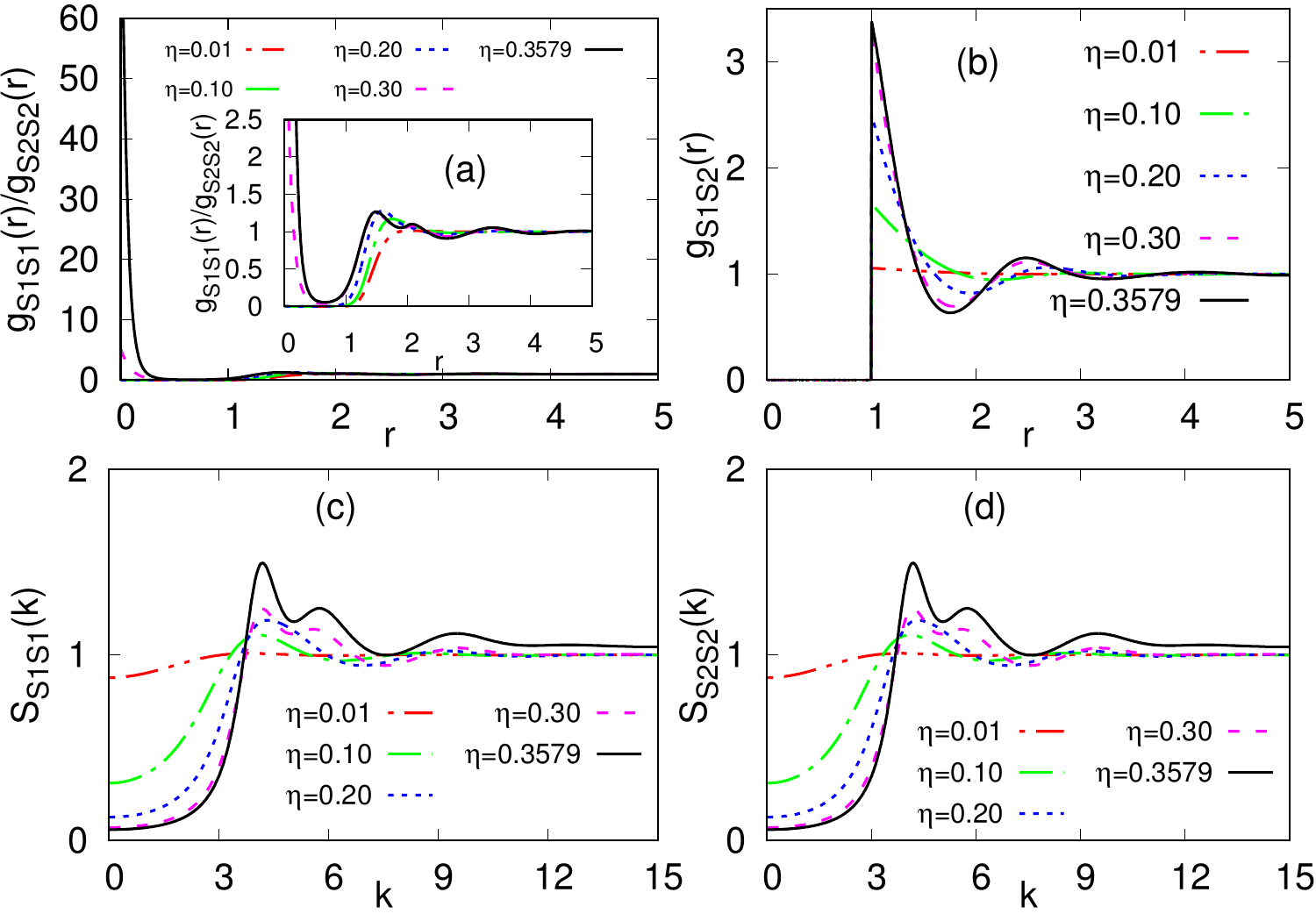}
		\caption{(a) Radial distribution functions $g(r)$ for like pairs ($S_1$--$S_1$ and $S_2$--$S_2$) in the Soft--Soft--Hard system at $m=3$, $\varepsilon=15$, and $\phi_{S2}=0.5$ for varying packing fraction $\eta$. (b) Radial distribution function for unlike pairs ($S_1$--$S_2$), showing hard-sphere exclusion between species. (c) Partial structure factor $S_{S_1S_1}(k)$. (d) Partial structure factor $S_{S_2S_2}(k)$. The absence of a dominant finite-$k$ peak is consistent with a dispersed state with local clustering of like species but no long-range compositional modulation.}
		\label{fig:fig9}
	\end{figure*}
	
	The radial distribution functions shown in Fig.~\ref{fig:fig9}(a) exhibit pronounced peaks at $r=0$ for both $S_1$--$S_1$ and $S_2$--$S_2$ pairs, indicating extensive overlap and the formation of clusters composed exclusively of like particles. Such clustering is a generic feature of systems governed by bounded repulsive interactions and has been widely reported in ultrasoft and penetrable-particle models~\cite{Likos_2001_E,Mladek_2006_PRL_ClusterPhases}. In contrast, the radial distribution function for unlike pairs (Fig.~\ref{fig:fig9}(b)) displays a sharp contact peak at $r=\sigma$ and vanishes for $r<\sigma$, confirming that the hard-sphere cross-interaction completely suppresses overlap between $S_1$ and $S_2$ particles.
	
	Despite this strong local structuring, the corresponding structure factors in Fig.~\ref{fig:fig9}(c,d) remain featureless across the entire range of packing fractions studied and do not exhibit peaks at finite wavevectors. The absence of a finite-$k$ instability indicates that neither microphase separation nor macrophase separation occurs, even though both components form clusters individually. This demonstrates that bounded self-interactions alone are insufficient to generate microphase separation when the cross interaction prevents mutual overlap and thereby suppresses coupling between local clustering and long-wavelength density fluctuations.
	
	\begin{figure}[tbp]
		\centering
		\includegraphics[width=\linewidth]{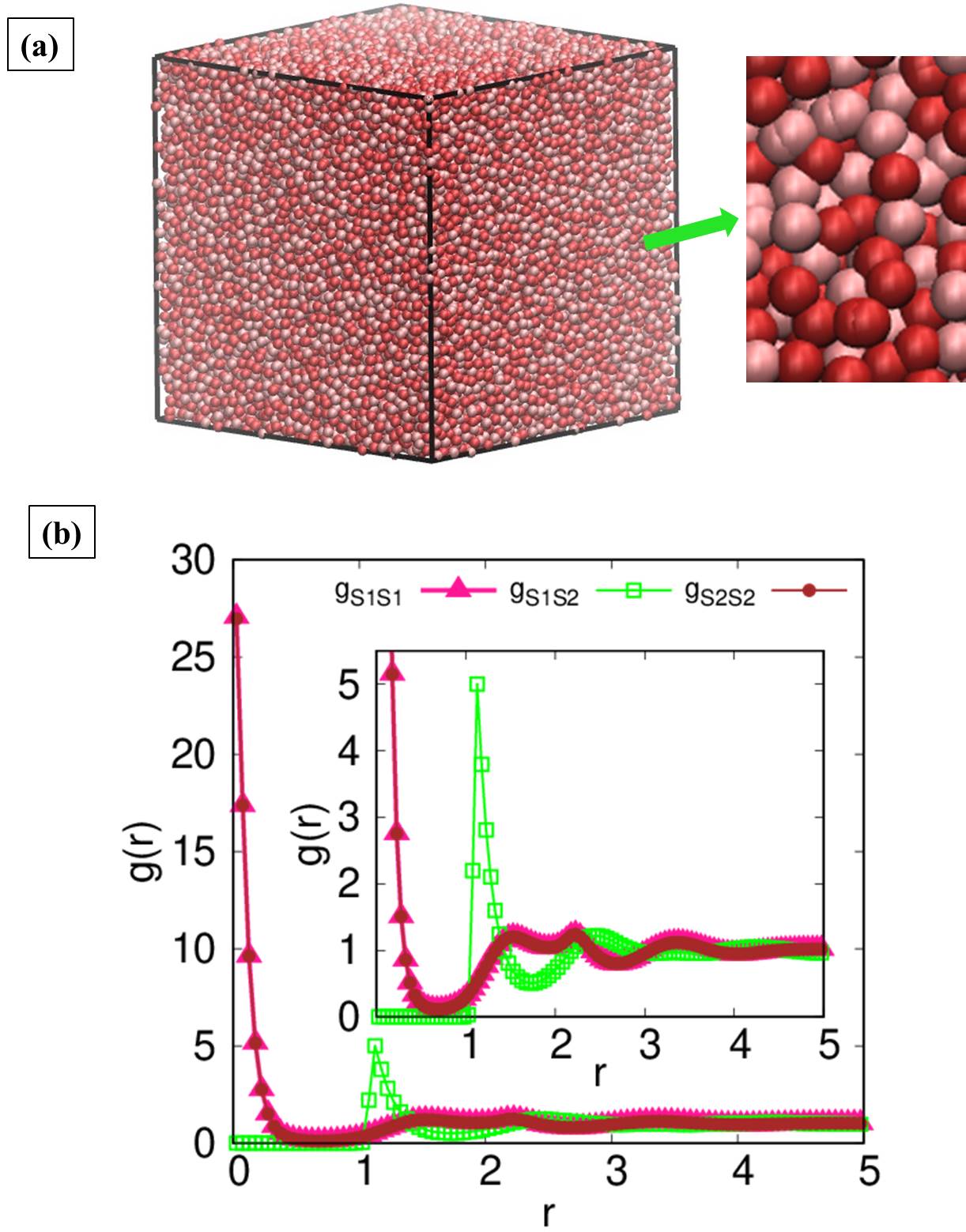}
		\caption{Molecular dynamics results for the Soft--Soft--Hard system at $m=3$, $\varepsilon = 15$, $\phi_{S2}=0.5$, and $\eta = 0.4$ in a simulation box of size 60. (a) Equilibrium snapshot showing a dispersed mixture with overlapping clusters of like particles. (b) Corresponding radial distribution functions.}
	\end{figure}
	
	Molecular dynamics simulations shown in Fig.~10 corroborate the theoretical predictions and provide a real-space perspective on the equilibrium morphology. The simulation snapshot (Fig.~10(a)) reveals that clusters of $S_1$ and $S_2$ particles form independently and remain spatially mixed throughout the simulation box, without developing extended compositional domains or periodic ordering. The corresponding radial distribution functions (Fig.~10(b)) display large peaks at $r=0$ for like pairs, consistent with extensive overlap and cluster formation, while the unlike-pair correlations exhibit a contact peak at $r=\sigma$ characteristic of hard-sphere exclusion. Beyond a few particle diameters, all correlation functions rapidly approach unity, indicating the absence of long-range density modulation.
	
The absence of microphase separation in the SSH system is consistent with earlier studies of binary mixtures of ultrasoft particles. Archer and co-workers showed that layered or modulated ordering in binary Gaussian core model fluids arises only when interactions between unlike species remain bounded and penetrable~\cite{Archer2004a, Archer2004lamda, Archer2007, Overduin_2008}. When mutual overlap between different species is prohibited, clustering remains confined within individual components and does not propagate into collective finite-wavelength ordering. Similarly, Overduin and Likos demonstrated that mixed cluster phases and compositional modulation in ultrasoft mixtures rely critically on the softness and range of the cross interaction~\cite{Overduin_2008}. The present results therefore reinforce the conclusion that penetrability of the \emph{cross interaction}, rather than bounded self-interactions alone, is the decisive ingredient for stabilizing microphase-separated states in binary soft-matter systems.

The results for the \emph{Hard--Soft--Hard} and \emph{Soft--Soft--Hard} systems further demonstrate that hard-sphere cross interactions fundamentally suppress microphase separation, even when one or both components exhibit strong local clustering due to bounded self-interactions. In both systems, clustering remains confined within individual species and does not propagate into finite-wavelength compositional ordering or large-scale segregation. The absence of pronounced finite-$k$ peaks in the structure factors, together with the homogeneous morphologies observed in molecular dynamics simulations, confirms that penetrability of the cross interaction is a necessary condition for coupling local clustering to collective, finite-wavelength instabilities. These findings establish that bounded self-interactions alone are insufficient to stabilize microphase-separated states and identify cross-interaction softness as the decisive control parameter governing the emergence of microphase separation in binary mixtures of hard and soft particles.

This conclusion naturally connects to recent studies of nearly hard-sphere systems, where introducing a \emph{minimal degree of softness} was shown to unlock otherwise inaccessible phase behavior. In particular, Wang \emph{et al.} demonstrated that slight deviations from perfect hard-sphere interactions enable access to the elusive fluid–crystal coexistence state in monodisperse systems through an entropy-exchange mechanism~\cite{wang2025elusive,wang2025frenkel}. Viewed in this broader context, the present results suggest that cross-interaction softness in multicomponent mixtures acts as a generalized and highly effective route for redistributing entropy and stabilizing complex collective ordering. While minimal softness in single-component systems facilitates coexistence between uniform phases, penetrable cross interactions in mixtures promote finite-wavelength compositional ordering, highlighting softness as a unifying design principle for accessing nontrivial phase behavior in colloidal and soft-matter systems.

\section*{Conclusion and Future Direction}

In this work, we investigated the role of cross interactions in governing the phase behavior of binary mixtures of hard and soft particles using a combination of coarse-grained modeling, integral-equation theory, and molecular dynamics simulations. By systematically comparing four representative systems—\emph{Hard--Soft--Soft}, \emph{Hard--Hard--Soft}, \emph{Hard--Soft--Hard}, and \emph{Soft--Soft--Hard}—we isolated the effect of cross-interaction character while keeping the self-interactions fixed. This approach allowed us to identify the key physical ingredients responsible for dispersed states, macrophase separation, and microphase separation in repulsively interacting soft-matter systems.

A central finding of this study is that \emph{penetrability of the cross interaction is the decisive control parameter for microphase separation}. In systems where unlike particles interact via bounded repulsive potentials, such as the \emph{Hard--Soft--Soft} and \emph{Hard--Hard--Soft} systems, partial overlap between dissimilar components suppresses complete demixing and stabilizes finite-wavelength compositional ordering. These systems exhibit rich phase behavior, including particle-scale ordering, macrophase separation at extreme compositions, and microphase separation at intermediate compositions. In contrast, when cross interactions are purely hard-sphere, as in the \emph{Hard--Soft--Hard} and \emph{Soft--Soft--Hard} systems, microphase separation is entirely suppressed, even though strong local clustering may occur within individual components.

These results stand in sharp contrast to the behavior of symmetric hard-sphere mixtures, which undergo freezing or glass transitions rather than microphase separation when all interactions are hard~\cite{Pusey_1985}. Our findings are instead consistent with earlier studies of bounded-potential mixtures, which have shown that microphase separation arises from the competition between excluded-volume effects and penetrable interactions~\cite{Likos_2001_E,Overduin_2008}. Importantly, the present work demonstrates that clustering within individual components is not sufficient to induce phase separation; rather, it is the ability of unlike particles to partially overlap that enables coupling between local clustering and long-wavelength density modulations.

Comparison between RISM theory and molecular dynamics simulations further reveals that while integral-equation approaches reliably predict the onset and topology of phase transitions, particle-resolved simulations capture additional features such as hierarchical ordering and multiscale densification near crossover regimes. In particular, for systems with bounded cross interactions, simulations reveal coexistence of local microphase ordering and large-scale structural rearrangements that are only partially captured by mean-field theories. Nevertheless, the overall qualitative agreement between theory and simulation confirms the robustness of the underlying physical picture. Taken together, the results of this work establish \emph{cross-interaction softness as a fundamental design principle} for controlling self-organization in binary colloidal and soft-matter systems. By tuning only the interactions between unlike particles, it is possible to switch between dispersed states, macrophase separation, and microphase separation without introducing attractive forces.

The present study focused on the generalized exponential model with a fixed steepness parameter ($m=3$) to isolate the effect of cross-interaction character. An important extension of this work is the systematic exploration of how the \emph{steepness of bounded potentials} influences phase behavior in binary mixtures. Previous studies have shown that GEM potentials exhibit re-entrant melting for $m \leq 2$ and clustering behavior for $m > 2$~\cite{Likos2001a,Coslovich2013}. Extending the present framework to mixtures interacting via GEM potentials with different steepness values, and over a wide range of interaction strengths, will enable a unified understanding of how softness, penetrability, and interaction range collectively control microphase and macrophase separation. Another promising direction is the quantitative characterization of \emph{ordering length scales} across theory and simulation. In this work, microphase separation length scales predicted by RISM theory were found to be systematically smaller than those observed in molecular dynamics simulations. Understanding the origin of this discrepancy—whether due to finite-size effects, fluctuation corrections beyond mean-field theory, or kinetic pathways—remains an open question and motivates further theoretical refinement.

Finally, the coarse-grained modeling strategy employed here naturally lends itself to applications in polymer nanocomposites, soft colloidal alloys, and multicomponent polymeric systems, where effective interactions between components can be tuned through chemistry or external fields. Extending the present approach to dynamically evolving systems, external confinement, or nonequilibrium driving may provide additional routes to programmable self-assembly and functional material design. These directions will be addressed in future work, building on the mechanistic insights established here regarding the central role of cross-interaction softness in soft-matter phase behavior.

\section*{Acknowledgments}
We would like to acknowledge the support of the C-DAC, IIT Bombay for the computational resources. 
	
\section*{Declaration of Competing Interests}
The authors declare that they have no known competing financial interests or personal relationships that could have appeared to influence the work reported in this paper.

\section*{Data Availability Statement}
That data that support the findings of this study are available from the corresponding author upon reasonable request.
	
\section{Appendix}
	
	\setcounter{figure}{0}
	\renewcommand{\thefigure}{A\arabic{figure}}
	\subsection{Robustness of multiscale ordering with respect to initial conditions}
	\begin{figure}[!htb]
		\centering
		\includegraphics[width=0.95\linewidth]{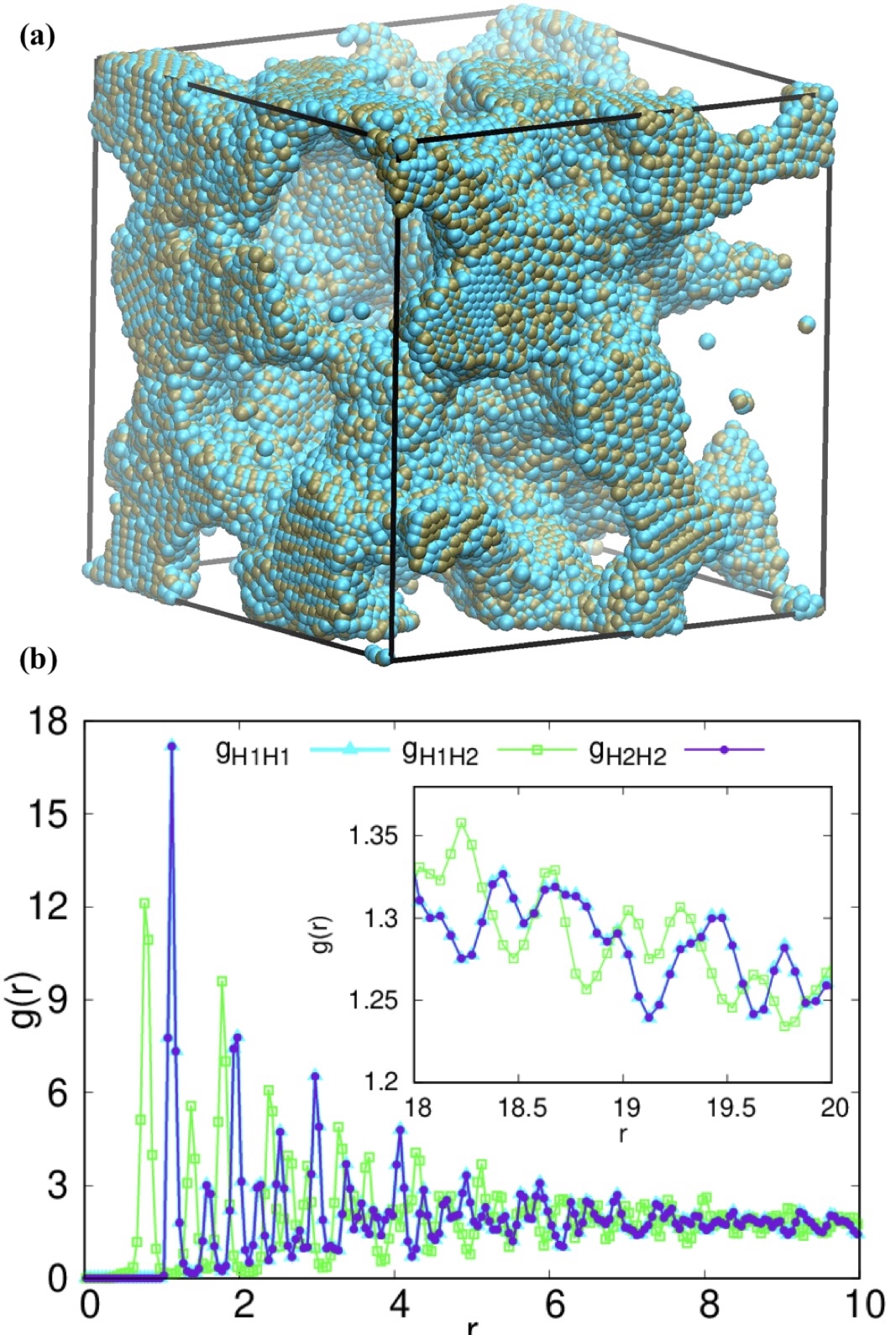}
		\caption{(a) MD snapshot for $m=3$, $\varepsilon/k_BT=15$, $\phi_{H_2}=0.5$, and $\eta=0.42$ with simulation box size 60, showing alternating $H_1$- and $H_2$-rich domains. (b) Corresponding radial distribution functions. The inset highlights short-range compositional ordering, while the slow decay of oscillations reflects long-range structural correlations.}
		\label{fig:figA1}
	\end{figure}
	
	\clearpage
	

\begin{thebibliography}{67}%
\makeatletter
\providecommand \@ifxundefined [1]{%
 \@ifx{#1\undefined}
}%
\providecommand \@ifnum [1]{%
 \ifnum #1\expandafter \@firstoftwo
 \else \expandafter \@secondoftwo
 \fi
}%
\providecommand \@ifx [1]{%
 \ifx #1\expandafter \@firstoftwo
 \else \expandafter \@secondoftwo
 \fi
}%
\providecommand \natexlab [1]{#1}%
\providecommand \enquote  [1]{``#1''}%
\providecommand \bibnamefont  [1]{#1}%
\providecommand \bibfnamefont [1]{#1}%
\providecommand \citenamefont [1]{#1}%
\providecommand \href@noop [0]{\@secondoftwo}%
\providecommand \href [0]{\begingroup \@sanitize@url \@href}%
\providecommand \@href[1]{\@@startlink{#1}\@@href}%
\providecommand \@@href[1]{\endgroup#1\@@endlink}%
\providecommand \@sanitize@url [0]{\catcode `\\12\catcode `\$12\catcode
  `\&12\catcode `\#12\catcode `\^12\catcode `\_12\catcode `\%12\relax}%
\providecommand \@@startlink[1]{}%
\providecommand \@@endlink[0]{}%
\providecommand \url  [0]{\begingroup\@sanitize@url \@url }%
\providecommand \@url [1]{\endgroup\@href {#1}{\urlprefix }}%
\providecommand \urlprefix  [0]{URL }%
\providecommand \Eprint [0]{\href }%
\providecommand \doibase [0]{https://doi.org/}%
\providecommand \selectlanguage [0]{\@gobble}%
\providecommand \bibinfo  [0]{\@secondoftwo}%
\providecommand \bibfield  [0]{\@secondoftwo}%
\providecommand \translation [1]{[#1]}%
\providecommand \BibitemOpen [0]{}%
\providecommand \bibitemStop [0]{}%
\providecommand \bibitemNoStop [0]{.\EOS\space}%
\providecommand \EOS [0]{\spacefactor3000\relax}%
\providecommand \BibitemShut  [1]{\csname bibitem#1\endcsname}%
\let\auto@bib@innerbib\@empty
\bibitem [{\citenamefont {Tadros}(2011)}]{Tadros_2011}%
  \BibitemOpen
  \bibinfo {editor} {\bibfnamefont {T.~F.}\ \bibnamefont {Tadros}},\ ed.,\
  \href@noop {} {\emph {\bibinfo {title} {Colloids in Paints}}}\ (\bibinfo
  {publisher} {Wiley-VCH},\ \bibinfo {year} {2011})\BibitemShut {NoStop}%
\bibitem [{\citenamefont {Tadros}(2013)}]{Tadros_2013}%
  \BibitemOpen
  \bibfield  {author} {\bibinfo {author} {\bibfnamefont {T.~F.}\ \bibnamefont
  {Tadros}},\ }\href@noop {} {\emph {\bibinfo {title} {Suspension Concentrates:
  Preparation, Stability and Industrial Applications}}}\ (\bibinfo  {publisher}
  {Wiley},\ \bibinfo {year} {2013})\BibitemShut {NoStop}%
\bibitem [{\citenamefont {Nappert}\ and\ \citenamefont
  {Reed}(1991)}]{Nappert_Reed_1991}%
  \BibitemOpen
  \bibfield  {author} {\bibinfo {author} {\bibfnamefont {D.~H.}\ \bibnamefont
  {Nappert}}\ and\ \bibinfo {author} {\bibfnamefont {R.~L.}\ \bibnamefont
  {Reed}},\ }\href@noop {} {\emph {\bibinfo {title} {Paints, Coatings and
  Solvents}}}\ (\bibinfo  {publisher} {Elsevier},\ \bibinfo {year}
  {1991})\BibitemShut {NoStop}%
\bibitem [{\citenamefont {Keddie}(1997)}]{Keddie_1997}%
  \BibitemOpen
  \bibfield  {author} {\bibinfo {author} {\bibfnamefont {J.~L.}\ \bibnamefont
  {Keddie}},\ }\bibfield  {title} {\enquote {\bibinfo {title} {Film formation
  of latex},}\ }\href@noop {} {\bibfield  {journal} {\bibinfo  {journal}
  {Materials Science and Engineering: R}\ }\textbf {\bibinfo {volume} {21}},\
  \bibinfo {pages} {101--170} (\bibinfo {year} {1997})}\BibitemShut {NoStop}%
\bibitem [{\citenamefont {Dickinson}(1992)}]{Dickinson_1992}%
  \BibitemOpen
  \bibfield  {author} {\bibinfo {author} {\bibfnamefont {E.}~\bibnamefont
  {Dickinson}},\ }\href@noop {} {\emph {\bibinfo {title} {An Introduction to
  Food Colloids}}}\ (\bibinfo  {publisher} {Oxford University Press},\ \bibinfo
  {year} {1992})\BibitemShut {NoStop}%
\bibitem [{\citenamefont {McClements}(2015)}]{McClements_2015}%
  \BibitemOpen
  \bibfield  {author} {\bibinfo {author} {\bibfnamefont {D.~J.}\ \bibnamefont
  {McClements}},\ }\href {https://doi.org/10.1201/b18868} {\emph {\bibinfo
  {title} {Food Emulsions: Principles, Practices, and Techniques}}},\ \bibinfo
  {edition} {3rd}\ ed.\ (\bibinfo  {publisher} {CRC Press},\ \bibinfo {year}
  {2015})\BibitemShut {NoStop}%
\bibitem [{\citenamefont {Tadros}(1992)}]{Tadros_1992}%
  \BibitemOpen
  \bibfield  {author} {\bibinfo {author} {\bibfnamefont {T.}~\bibnamefont
  {Tadros}},\ }\bibfield  {title} {\enquote {\bibinfo {title} {Industrial
  applications of dispersions},}\ }\href@noop {} {\bibfield  {journal}
  {\bibinfo  {journal} {Advances in Colloid and Interface Science}\ } (\bibinfo
  {year} {1992})}\BibitemShut {NoStop}%
\bibitem [{\citenamefont {M{\"u}ller}\ and\ \citenamefont
  {Peters}(1998)}]{MullerPeters_1998}%
  \BibitemOpen
  \bibfield  {author} {\bibinfo {author} {\bibfnamefont {R.~H.}\ \bibnamefont
  {M{\"u}ller}}\ and\ \bibinfo {author} {\bibfnamefont {K.}~\bibnamefont
  {Peters}},\ }\bibfield  {title} {\enquote {\bibinfo {title} {Nanosuspensions
  for the formulation of poorly soluble drugs},}\ }\href
  {https://doi.org/10.1016/S0378-5173(97)00311-6} {\bibfield  {journal}
  {\bibinfo  {journal} {International Journal of Pharmaceutics}\ }\textbf
  {\bibinfo {volume} {160}},\ \bibinfo {pages} {229--237} (\bibinfo {year}
  {1998})}\BibitemShut {NoStop}%
\bibitem [{\citenamefont {Navascués}, \citenamefont {Velasco},\ and\
  \citenamefont {Mederos}(2004)}]{Mederos_2004}%
  \BibitemOpen
  \bibfield  {author} {\bibinfo {author} {\bibfnamefont {G.}~\bibnamefont
  {Navascués}}, \bibinfo {author} {\bibfnamefont {E.}~\bibnamefont
  {Velasco}},\ and\ \bibinfo {author} {\bibfnamefont {L.}~\bibnamefont
  {Mederos}},\ }\bibfield  {title} {\enquote {\bibinfo {title} {Theories of
  phase behavior of colloidal suspensions},}\ }\href@noop {} {\bibfield
  {journal} {\bibinfo  {journal} {International Journal of Modern Physics B}\ }
  (\bibinfo {year} {2004})}\BibitemShut {NoStop}%
\bibitem [{\citenamefont {Tadros}(2015)}]{Tadros_2015}%
  \BibitemOpen
  \bibfield  {author} {\bibinfo {author} {\bibfnamefont {T.~F.}\ \bibnamefont
  {Tadros}},\ }\href@noop {} {\emph {\bibinfo {title} {Nanodispersions}}}\
  (\bibinfo  {publisher} {De Gruyter},\ \bibinfo {year} {2015})\BibitemShut
  {NoStop}%
\bibitem [{\citenamefont {Likos}(2001)}]{Likos_2001_E}%
  \BibitemOpen
  \bibfield  {author} {\bibinfo {author} {\bibfnamefont {C.~N.}\ \bibnamefont
  {Likos}},\ }\bibfield  {title} {\enquote {\bibinfo {title} {Effective
  interactions in soft condensed matter physics},}\ }\href@noop {} {\bibfield
  {journal} {\bibinfo  {journal} {Phys. Rep.}\ }\textbf {\bibinfo {volume}
  {348}},\ \bibinfo {pages} {267--439} (\bibinfo {year} {2001})}\BibitemShut
  {NoStop}%
\bibitem [{\citenamefont {Gasser}(2009)}]{Gasser_2009}%
  \BibitemOpen
  \bibfield  {author} {\bibinfo {author} {\bibfnamefont {U.}~\bibnamefont
  {Gasser}},\ }\bibfield  {title} {\enquote {\bibinfo {title} {Crystallization
  in three- and two-dimensional colloidal suspensions},}\ }\href@noop {}
  {\bibfield  {journal} {\bibinfo  {journal} {Journal of Physics Condensed
  Matter}\ } (\bibinfo {year} {2009})}\BibitemShut {NoStop}%
\bibitem [{\citenamefont {Sciortino}\ and\ \citenamefont
  {Tartaglia}(2005)}]{Sciortino_Tartaglia_2005}%
  \BibitemOpen
  \bibfield  {author} {\bibinfo {author} {\bibfnamefont {F.}~\bibnamefont
  {Sciortino}}\ and\ \bibinfo {author} {\bibfnamefont {P.}~\bibnamefont
  {Tartaglia}},\ }\bibfield  {title} {\enquote {\bibinfo {title} {Glassy
  colloidal systems},}\ }\href@noop {} {\bibfield  {journal} {\bibinfo
  {journal} {Advances In Physics}\ } (\bibinfo {year} {2005})}\BibitemShut
  {NoStop}%
\bibitem [{\citenamefont {Pusey}\ and\ \citenamefont
  {Megen}(1985)}]{Pusey_1985}%
  \BibitemOpen
  \bibfield  {author} {\bibinfo {author} {\bibfnamefont {P.}~\bibnamefont
  {Pusey}}\ and\ \bibinfo {author} {\bibfnamefont {W.~V.}\ \bibnamefont
  {Megen}},\ }\bibfield  {title} {\enquote {\bibinfo {title} {Phase behaviour
  of concentrated suspensions of nearly hard colloidal spheres},}\ }\href@noop
  {} {\bibfield  {journal} {\bibinfo  {journal} {Nature}\ } (\bibinfo {year}
  {1985})}\BibitemShut {NoStop}%
\bibitem [{\citenamefont {Schätzel}\ and\ \citenamefont
  {Ackerson}(1992)}]{Ackerson_1992}%
  \BibitemOpen
  \bibfield  {author} {\bibinfo {author} {\bibfnamefont {K.}~\bibnamefont
  {Schätzel}}\ and\ \bibinfo {author} {\bibfnamefont {B.}~\bibnamefont
  {Ackerson}},\ }\bibfield  {title} {\enquote {\bibinfo {title}
  {Crystallization of hard sphere colloids},}\ }\href@noop {} {\bibfield
  {journal} {\bibinfo  {journal} {Physica Scripta}\ } (\bibinfo {year}
  {1992})}\BibitemShut {NoStop}%
\bibitem [{\citenamefont {Degiorgio}, \citenamefont {Piazza},\ and\
  \citenamefont {Jones}(1995)}]{Degiorgio_1995}%
  \BibitemOpen
  \bibfield  {author} {\bibinfo {author} {\bibfnamefont {V.}~\bibnamefont
  {Degiorgio}}, \bibinfo {author} {\bibfnamefont {R.}~\bibnamefont {Piazza}},\
  and\ \bibinfo {author} {\bibfnamefont {R.~B.}\ \bibnamefont {Jones}},\
  }\bibfield  {title} {\enquote {\bibinfo {title} {Rotational diffusion in
  concentrated colloidal dispersions of hard spheres},}\ }\href@noop {}
  {\bibfield  {journal} {\bibinfo  {journal} {Physical review. E, Statistical
  physics, plasmas, fluids, and related interdisciplinary topics}\ } (\bibinfo
  {year} {1995})}\BibitemShut {NoStop}%
\bibitem [{\citenamefont {Ilett}\ \emph {et~al.}(1995)\citenamefont {Ilett},
  \citenamefont {Orrock}, \citenamefont {Poon},\ and\ \citenamefont
  {Pusey}}]{Pusey_1995}%
  \BibitemOpen
  \bibfield  {author} {\bibinfo {author} {\bibfnamefont {S.~M.}\ \bibnamefont
  {Ilett}}, \bibinfo {author} {\bibfnamefont {A.}~\bibnamefont {Orrock}},
  \bibinfo {author} {\bibfnamefont {W.~C.~K.}\ \bibnamefont {Poon}},\ and\
  \bibinfo {author} {\bibfnamefont {P.~N.}\ \bibnamefont {Pusey}},\ }\bibfield
  {title} {\enquote {\bibinfo {title} {Phase behavior of a model
  colloid-polymer mixture},}\ }\href@noop {} {\bibfield  {journal} {\bibinfo
  {journal} {Phys. Rev. E}\ }\textbf {\bibinfo {volume} {51}},\ \bibinfo
  {pages} {1344--1352} (\bibinfo {year} {1995})}\BibitemShut {NoStop}%
\bibitem [{\citenamefont {Vlassopoulos}, \citenamefont {Stiakakis},\ and\
  \citenamefont {Kapnistos}(2007)}]{vlassopoulos2007}%
  \BibitemOpen
  \bibfield  {author} {\bibinfo {author} {\bibfnamefont {D.}~\bibnamefont
  {Vlassopoulos}}, \bibinfo {author} {\bibfnamefont {E.}~\bibnamefont
  {Stiakakis}},\ and\ \bibinfo {author} {\bibfnamefont {M.}~\bibnamefont
  {Kapnistos}},\ }\bibfield  {title} {\enquote {\bibinfo {title} {Model soft
  colloids out of equilibrium: glass-like and re-entrant transitions},}\
  }\href@noop {} {\bibfield  {journal} {\bibinfo  {journal} {Rheol. Rev}\ ,\
  \bibinfo {pages} {179--252}} (\bibinfo {year} {2007})}\BibitemShut {NoStop}%
\bibitem [{\citenamefont {Overduin}\ and\ \citenamefont
  {Likos}(2008)}]{Overduin_2008}%
  \BibitemOpen
  \bibfield  {author} {\bibinfo {author} {\bibfnamefont {S.}~\bibnamefont
  {Overduin}}\ and\ \bibinfo {author} {\bibfnamefont {C.}~\bibnamefont
  {Likos}},\ }\bibfield  {title} {\enquote {\bibinfo {title} {Phase behaviour
  in binary mixtures of ultrasoft repulsive particles},}\ }\href@noop {}
  {\bibfield  {journal} {\bibinfo  {journal} {Europhysics Letters}\ } (\bibinfo
  {year} {2008})}\BibitemShut {NoStop}%
\bibitem [{\citenamefont {Likos}, \citenamefont {Hoffmann},\ and\ \citenamefont
  {Löwen}(2001)}]{Likos_2001}%
  \BibitemOpen
  \bibfield  {author} {\bibinfo {author} {\bibfnamefont {C.}~\bibnamefont
  {Likos}}, \bibinfo {author} {\bibfnamefont {N.}~\bibnamefont {Hoffmann}},\
  and\ \bibinfo {author} {\bibfnamefont {H.}~\bibnamefont {Löwen}},\
  }\bibfield  {title} {\enquote {\bibinfo {title} {Exotic fluids and crystals
  of soft polymeric colloids},}\ }\href@noop {} {\  (\bibinfo {year}
  {2001})}\BibitemShut {NoStop}%
\bibitem [{\citenamefont {Curk}\ \emph {et~al.}(2012)\citenamefont {Curk},
  \citenamefont {de~Hoogh}, \citenamefont {Martinez-Veracoechea}, \citenamefont
  {Eiser}, \citenamefont {Frenkel}, \citenamefont {Dobnikar},\ and\
  \citenamefont {Leunissen}}]{Curk_2012}%
  \BibitemOpen
  \bibfield  {author} {\bibinfo {author} {\bibfnamefont {T.}~\bibnamefont
  {Curk}}, \bibinfo {author} {\bibfnamefont {A.}~\bibnamefont {de~Hoogh}},
  \bibinfo {author} {\bibfnamefont {F.~J.}\ \bibnamefont
  {Martinez-Veracoechea}}, \bibinfo {author} {\bibfnamefont {E.}~\bibnamefont
  {Eiser}}, \bibinfo {author} {\bibfnamefont {D.}~\bibnamefont {Frenkel}},
  \bibinfo {author} {\bibfnamefont {J.}~\bibnamefont {Dobnikar}},\ and\
  \bibinfo {author} {\bibfnamefont {M.~E.}\ \bibnamefont {Leunissen}},\
  }\bibfield  {title} {\enquote {\bibinfo {title} {Layering, freezing, and
  re-entrant melting of hard spheres in soft confinement},}\ }\href@noop {}
  {\bibfield  {journal} {\bibinfo  {journal} {Phys. Rev. E}\ }\textbf {\bibinfo
  {volume} {85}},\ \bibinfo {pages} {021502} (\bibinfo {year}
  {2012})}\BibitemShut {NoStop}%
\bibitem [{\citenamefont {Zhu}\ and\ \citenamefont {Lu}(2010)}]{Zhu_Lu_2010}%
  \BibitemOpen
  \bibfield  {author} {\bibinfo {author} {\bibfnamefont {Y.}~\bibnamefont
  {Zhu}}\ and\ \bibinfo {author} {\bibfnamefont {Z.}~\bibnamefont {Lu}},\
  }\bibfield  {title} {\enquote {\bibinfo {title} {Phase diagram of spherical
  particles interacted with harmonic repulsions},}\ }\href@noop {} {\bibfield
  {journal} {\bibinfo  {journal} {The Journal of Chemical Physics}\ } (\bibinfo
  {year} {2010})}\BibitemShut {NoStop}%
\bibitem [{\citenamefont {Kalyuzhnyi}\ \emph {et~al.}(2024)\citenamefont
  {Kalyuzhnyi}, \citenamefont {Patsahan}, \citenamefont {Holovko},\ and\
  \citenamefont {Cummings}}]{Kalyuzhnyi_2024}%
  \BibitemOpen
  \bibfield  {author} {\bibinfo {author} {\bibfnamefont {Y.}~\bibnamefont
  {Kalyuzhnyi}}, \bibinfo {author} {\bibfnamefont {T.}~\bibnamefont
  {Patsahan}}, \bibinfo {author} {\bibfnamefont {M.}~\bibnamefont {Holovko}},\
  and\ \bibinfo {author} {\bibfnamefont {P.~T.}\ \bibnamefont {Cummings}},\
  }\bibfield  {title} {\enquote {\bibinfo {title} {Correction: Phase behavior
  of patchy colloids confined in patchy porous media.}}\ }\href@noop {}
  {\bibfield  {journal} {\bibinfo  {journal} {Nanoscale}\ } (\bibinfo {year}
  {2024})}\BibitemShut {NoStop}%
\bibitem [{\citenamefont {Lang}, \citenamefont {Likos},\ and\ \citenamefont
  {Watzlawek}(1999)}]{Lang_1999}%
  \BibitemOpen
  \bibfield  {author} {\bibinfo {author} {\bibfnamefont {A.}~\bibnamefont
  {Lang}}, \bibinfo {author} {\bibfnamefont {C.}~\bibnamefont {Likos}},\ and\
  \bibinfo {author} {\bibfnamefont {M.}~\bibnamefont {Watzlawek}},\ }\bibfield
  {title} {\enquote {\bibinfo {title} {Fluid andsolid phases of the gaussian
  core model},}\ }\href@noop {} {\  (\bibinfo {year} {1999})}\BibitemShut
  {NoStop}%
\bibitem [{\citenamefont {Travis}\ and\ \citenamefont
  {Sadus}(2024)}]{Travis_2024}%
  \BibitemOpen
  \bibfield  {author} {\bibinfo {author} {\bibfnamefont {K.~P.}\ \bibnamefont
  {Travis}}\ and\ \bibinfo {author} {\bibfnamefont {R.~J.}\ \bibnamefont
  {Sadus}},\ }\bibfield  {title} {\enquote {\bibinfo {title} {Solid–fluid
  equilibria of atoms with soft repulsive and short-range cohesive
  interactions},}\ }\href@noop {} {\bibfield  {journal} {\bibinfo  {journal}
  {The Journal of Physical Chemistry B}\ }\textbf {\bibinfo {volume} {128}},\
  \bibinfo {pages} {2922--2929} (\bibinfo {year} {2024})}\BibitemShut {NoStop}%
\bibitem [{\citenamefont {Kanai}\ \emph {et~al.}(2015)\citenamefont {Kanai},
  \citenamefont {Boon}, \citenamefont {Lu}, \citenamefont {Sloutskin},
  \citenamefont {Schofield}, \citenamefont {Smallenburg}, \citenamefont {van
  Roij}, \citenamefont {Dijkstra},\ and\ \citenamefont {Weitz}}]{Kanai_2015}%
  \BibitemOpen
  \bibfield  {author} {\bibinfo {author} {\bibfnamefont {T.}~\bibnamefont
  {Kanai}}, \bibinfo {author} {\bibfnamefont {N.}~\bibnamefont {Boon}},
  \bibinfo {author} {\bibfnamefont {P.~J.}\ \bibnamefont {Lu}}, \bibinfo
  {author} {\bibfnamefont {E.}~\bibnamefont {Sloutskin}}, \bibinfo {author}
  {\bibfnamefont {A.~B.}\ \bibnamefont {Schofield}}, \bibinfo {author}
  {\bibfnamefont {F.}~\bibnamefont {Smallenburg}}, \bibinfo {author}
  {\bibfnamefont {R.}~\bibnamefont {van Roij}}, \bibinfo {author}
  {\bibfnamefont {M.}~\bibnamefont {Dijkstra}},\ and\ \bibinfo {author}
  {\bibfnamefont {D.~A.}\ \bibnamefont {Weitz}},\ }\bibfield  {title} {\enquote
  {\bibinfo {title} {Crystallization and reentrant melting of charged colloids
  in nonpolar solvents},}\ }\href@noop {} {\bibfield  {journal} {\bibinfo
  {journal} {Phys. Rev. E}\ }\textbf {\bibinfo {volume} {91}},\ \bibinfo
  {pages} {030301} (\bibinfo {year} {2015})}\BibitemShut {NoStop}%
\bibitem [{\citenamefont {Everts}, \citenamefont {Boon},\ and\ \citenamefont
  {Roij}(2015)}]{Everts_2015}%
  \BibitemOpen
  \bibfield  {author} {\bibinfo {author} {\bibfnamefont {J.}~\bibnamefont
  {Everts}}, \bibinfo {author} {\bibfnamefont {N.}~\bibnamefont {Boon}},\ and\
  \bibinfo {author} {\bibfnamefont {R.~V.}\ \bibnamefont {Roij}},\ }\bibfield
  {title} {\enquote {\bibinfo {title} {Density-induced reentrant melting of
  colloidal wigner crystals.}}\ }\href@noop {} {\bibfield  {journal} {\bibinfo
  {journal} {Physical chemistry chemical physics}\ } (\bibinfo {year}
  {2015})}\BibitemShut {NoStop}%
\bibitem [{\citenamefont {Götze}, \citenamefont {Archer},\ and\ \citenamefont
  {Likos}(2005)}]{Likos_2005}%
  \BibitemOpen
  \bibfield  {author} {\bibinfo {author} {\bibfnamefont {I.~O.}\ \bibnamefont
  {Götze}}, \bibinfo {author} {\bibfnamefont {A.~J.}\ \bibnamefont {Archer}},\
  and\ \bibinfo {author} {\bibfnamefont {C.~N.}\ \bibnamefont {Likos}},\
  }\bibfield  {title} {\enquote {\bibinfo {title} {Structure, phase behavior,
  and inhomogeneous fluid properties of binary dendrimer mixtures.}}\
  }\href@noop {} {\bibfield  {journal} {\bibinfo  {journal} {The Journal of
  chemical physics}\ } (\bibinfo {year} {2005})}\BibitemShut {NoStop}%
\bibitem [{\citenamefont {Bates}\ and\ \citenamefont
  {Fredrickson}(1999)}]{BatesFredrickson_1999}%
  \BibitemOpen
  \bibfield  {author} {\bibinfo {author} {\bibfnamefont {F.~S.}\ \bibnamefont
  {Bates}}\ and\ \bibinfo {author} {\bibfnamefont {G.~H.}\ \bibnamefont
  {Fredrickson}},\ }\bibfield  {title} {\enquote {\bibinfo {title} {Block
  copolymers---designer soft materials},}\ }\href
  {https://doi.org/10.1063/1.882522} {\bibfield  {journal} {\bibinfo  {journal}
  {Physics Today}\ }\textbf {\bibinfo {volume} {52}},\ \bibinfo {pages}
  {32--38} (\bibinfo {year} {1999})}\BibitemShut {NoStop}%
\bibitem [{\citenamefont {Carta}\ \emph {et~al.}(2011)\citenamefont {Carta},
  \citenamefont {Pini}, \citenamefont {Parola},\ and\ \citenamefont
  {Reatto}}]{Carta_2011}%
  \BibitemOpen
  \bibfield  {author} {\bibinfo {author} {\bibfnamefont {M.}~\bibnamefont
  {Carta}}, \bibinfo {author} {\bibfnamefont {D.}~\bibnamefont {Pini}},
  \bibinfo {author} {\bibfnamefont {A.}~\bibnamefont {Parola}},\ and\ \bibinfo
  {author} {\bibfnamefont {L.}~\bibnamefont {Reatto}},\ }\bibfield  {title}
  {\enquote {\bibinfo {title} {A density-functional theory study of microphase
  formation in binary gaussian mixtures},}\ }\href@noop {} {\  (\bibinfo {year}
  {2011})}\BibitemShut {NoStop}%
\bibitem [{\citenamefont {Discher}\ and\ \citenamefont
  {Eisenberg}(2002)}]{DischerEisenberg_2002}%
  \BibitemOpen
  \bibfield  {author} {\bibinfo {author} {\bibfnamefont {D.~E.}\ \bibnamefont
  {Discher}}\ and\ \bibinfo {author} {\bibfnamefont {A.}~\bibnamefont
  {Eisenberg}},\ }\bibfield  {title} {\enquote {\bibinfo {title} {Polymer
  vesicles},}\ }\href {https://doi.org/10.1126/science.1074972} {\bibfield
  {journal} {\bibinfo  {journal} {Science}\ }\textbf {\bibinfo {volume}
  {297}},\ \bibinfo {pages} {967--973} (\bibinfo {year} {2002})}\BibitemShut
  {NoStop}%
\bibitem [{\citenamefont {Ahmed}\ \emph {et~al.}(2006)\citenamefont {Ahmed},
  \citenamefont {Pakunlu}, \citenamefont {Brannan}, \citenamefont {Bates},
  \citenamefont {Minko},\ and\ \citenamefont {Discher}}]{Ahmed_2006}%
  \BibitemOpen
  \bibfield  {author} {\bibinfo {author} {\bibfnamefont {F.}~\bibnamefont
  {Ahmed}}, \bibinfo {author} {\bibfnamefont {R.~I.}\ \bibnamefont {Pakunlu}},
  \bibinfo {author} {\bibfnamefont {A.}~\bibnamefont {Brannan}}, \bibinfo
  {author} {\bibfnamefont {F.~S.}\ \bibnamefont {Bates}}, \bibinfo {author}
  {\bibfnamefont {T.}~\bibnamefont {Minko}},\ and\ \bibinfo {author}
  {\bibfnamefont {D.~E.}\ \bibnamefont {Discher}},\ }\bibfield  {title}
  {\enquote {\bibinfo {title} {Biodegradable polymersomes loaded with both
  paclitaxel and doxorubicin permeate and shrink tumors, inducing apoptosis in
  proportion to accumulated drug},}\ }\href
  {https://doi.org/10.1016/j.jconrel.2006.07.013} {\bibfield  {journal}
  {\bibinfo  {journal} {Journal of Controlled Release}\ }\textbf {\bibinfo
  {volume} {116}},\ \bibinfo {pages} {150--158} (\bibinfo {year}
  {2006})}\BibitemShut {NoStop}%
\bibitem [{\citenamefont {Discher}\ \emph {et~al.}(2007)\citenamefont
  {Discher}, \citenamefont {Ortiz}, \citenamefont {Srinivas}, \citenamefont
  {Klein}, \citenamefont {Kim}, \citenamefont {David}, \citenamefont {Cai},
  \citenamefont {Photos},\ and\ \citenamefont {Ahmed}}]{Discher_2007}%
  \BibitemOpen
  \bibfield  {author} {\bibinfo {author} {\bibfnamefont {D.~E.}\ \bibnamefont
  {Discher}}, \bibinfo {author} {\bibfnamefont {V.}~\bibnamefont {Ortiz}},
  \bibinfo {author} {\bibfnamefont {G.}~\bibnamefont {Srinivas}}, \bibinfo
  {author} {\bibfnamefont {M.~L.}\ \bibnamefont {Klein}}, \bibinfo {author}
  {\bibfnamefont {Y.}~\bibnamefont {Kim}}, \bibinfo {author} {\bibfnamefont
  {C.~A.}\ \bibnamefont {David}}, \bibinfo {author} {\bibfnamefont
  {S.}~\bibnamefont {Cai}}, \bibinfo {author} {\bibfnamefont {P.}~\bibnamefont
  {Photos}},\ and\ \bibinfo {author} {\bibfnamefont {F.}~\bibnamefont
  {Ahmed}},\ }\bibfield  {title} {\enquote {\bibinfo {title} {Emerging
  applications of polymersomes in delivery: From molecular dynamics to
  shrinkage of tumors},}\ }\href {https://doi.org/10.1002/0471781276.ch1}
  {\bibfield  {journal} {\bibinfo  {journal} {Advances in Chemical Physics}\
  }\textbf {\bibinfo {volume} {128}},\ \bibinfo {pages} {1--48} (\bibinfo
  {year} {2007})}\BibitemShut {NoStop}%
\bibitem [{\citenamefont {Nunes}\ and\ \citenamefont
  {Peinemann}(2006)}]{NunesPeinemann_2006}%
  \BibitemOpen
  \bibfield  {author} {\bibinfo {author} {\bibfnamefont {S.~P.}\ \bibnamefont
  {Nunes}}\ and\ \bibinfo {author} {\bibfnamefont {K.-V.}\ \bibnamefont
  {Peinemann}},\ }\bibfield  {title} {\enquote {\bibinfo {title} {Membranes for
  ultrafiltration and microfiltration from block copolymers},}\ }\href
  {https://doi.org/10.1002/adma.200600530} {\bibfield  {journal} {\bibinfo
  {journal} {Advanced Materials}\ }\textbf {\bibinfo {volume} {18}},\ \bibinfo
  {pages} {2519--2523} (\bibinfo {year} {2006})}\BibitemShut {NoStop}%
\bibitem [{\citenamefont {Nunes}\ \emph {et~al.}(2007)\citenamefont {Nunes},
  \citenamefont {Sougrat}, \citenamefont {Hooghan}, \citenamefont {Anjum},
  \citenamefont {Behzad}, \citenamefont {Pradeep},\ and\ \citenamefont
  {Peinemann}}]{Peinemann_2007}%
  \BibitemOpen
  \bibfield  {author} {\bibinfo {author} {\bibfnamefont {S.~P.}\ \bibnamefont
  {Nunes}}, \bibinfo {author} {\bibfnamefont {R.}~\bibnamefont {Sougrat}},
  \bibinfo {author} {\bibfnamefont {B.}~\bibnamefont {Hooghan}}, \bibinfo
  {author} {\bibfnamefont {D.~H.}\ \bibnamefont {Anjum}}, \bibinfo {author}
  {\bibfnamefont {A.~R.}\ \bibnamefont {Behzad}}, \bibinfo {author}
  {\bibfnamefont {N.}~\bibnamefont {Pradeep}},\ and\ \bibinfo {author}
  {\bibfnamefont {K.-V.}\ \bibnamefont {Peinemann}},\ }\bibfield  {title}
  {\enquote {\bibinfo {title} {Asymmetric superstructure formed in a block
  copolymer via phase separation},}\ }\href {https://doi.org/10.1038/nmat2038}
  {\bibfield  {journal} {\bibinfo  {journal} {Nature Materials}\ }\textbf
  {\bibinfo {volume} {6}},\ \bibinfo {pages} {609--614} (\bibinfo {year}
  {2007})}\BibitemShut {NoStop}%
\bibitem [{\citenamefont {Lang}\ \emph {et~al.}(2021)\citenamefont {Lang},
  \citenamefont {Arges}, \citenamefont {Kumar},\ and\ \citenamefont
  {Hillhouse}}]{LangKumarHillhouse_2021}%
  \BibitemOpen
  \bibfield  {author} {\bibinfo {author} {\bibfnamefont {C.}~\bibnamefont
  {Lang}}, \bibinfo {author} {\bibfnamefont {C.~G.}\ \bibnamefont {Arges}},
  \bibinfo {author} {\bibfnamefont {M.}~\bibnamefont {Kumar}},\ and\ \bibinfo
  {author} {\bibfnamefont {H.~W.}\ \bibnamefont {Hillhouse}},\ }\bibfield
  {title} {\enquote {\bibinfo {title} {Current status and future directions of
  self-assembled block copolymer membranes},}\ }\href
  {https://doi.org/10.1039/D1SM01368H} {\bibfield  {journal} {\bibinfo
  {journal} {Soft Matter}\ }\textbf {\bibinfo {volume} {17}},\ \bibinfo {pages}
  {7172--7192} (\bibinfo {year} {2021})}\BibitemShut {NoStop}%
\bibitem [{\citenamefont {Park}\ \emph {et~al.}(2003)\citenamefont {Park},
  \citenamefont {Harrison}, \citenamefont {Chaikin}, \citenamefont {Register},\
  and\ \citenamefont {Adamson}}]{Park_2003}%
  \BibitemOpen
  \bibfield  {author} {\bibinfo {author} {\bibfnamefont {M.}~\bibnamefont
  {Park}}, \bibinfo {author} {\bibfnamefont {C.}~\bibnamefont {Harrison}},
  \bibinfo {author} {\bibfnamefont {P.~M.}\ \bibnamefont {Chaikin}}, \bibinfo
  {author} {\bibfnamefont {R.~A.}\ \bibnamefont {Register}},\ and\ \bibinfo
  {author} {\bibfnamefont {D.~H.}\ \bibnamefont {Adamson}},\ }\bibfield
  {title} {\enquote {\bibinfo {title} {Block copolymer lithography: Periodic
  arrays of $\sim 10^{11}$ holes in 1 square centimeter},}\ }\href
  {https://doi.org/10.1126/science.1080719} {\bibfield  {journal} {\bibinfo
  {journal} {Science}\ }\textbf {\bibinfo {volume} {299}},\ \bibinfo {pages}
  {1716--1719} (\bibinfo {year} {2003})}\BibitemShut {NoStop}%
\bibitem [{\citenamefont {Albrecht}\ \emph {et~al.}(1997)\citenamefont
  {Albrecht} \emph {et~al.}}]{Albrecht_1997}%
  \BibitemOpen
  \bibfield  {author} {\bibinfo {author} {\bibfnamefont {T.~R.}\ \bibnamefont
  {Albrecht}} \emph {et~al.},\ }\bibfield  {title} {\enquote {\bibinfo {title}
  {Block copolymer lithography: Periodic arrays of $\sim 10^{11}$ holes in 1
  square centimeter},}\ }\href {https://doi.org/10.1126/science.276.5317.1401}
  {\bibfield  {journal} {\bibinfo  {journal} {Science}\ }\textbf {\bibinfo
  {volume} {276}},\ \bibinfo {pages} {1401--1404} (\bibinfo {year}
  {1997})}\BibitemShut {NoStop}%
\bibitem [{\citenamefont {Stoykovich}\ and\ \citenamefont
  {Nealey}(2006)}]{StoykovichNealey_2006}%
  \BibitemOpen
  \bibfield  {author} {\bibinfo {author} {\bibfnamefont {M.~P.}\ \bibnamefont
  {Stoykovich}}\ and\ \bibinfo {author} {\bibfnamefont {P.~F.}\ \bibnamefont
  {Nealey}},\ }\bibfield  {title} {\enquote {\bibinfo {title} {Block copolymers
  and conventional lithography},}\ }\href
  {https://doi.org/10.1016/S1369-7021(06)71619-4} {\bibfield  {journal}
  {\bibinfo  {journal} {Materials Today}\ }\textbf {\bibinfo {volume} {9}},\
  \bibinfo {pages} {20--29} (\bibinfo {year} {2006})}\BibitemShut {NoStop}%
\bibitem [{\citenamefont {Leibler}(1980)}]{Leibler_1980}%
  \BibitemOpen
  \bibfield  {author} {\bibinfo {author} {\bibfnamefont {L.}~\bibnamefont
  {Leibler}},\ }\bibfield  {title} {\enquote {\bibinfo {title} {Theory of
  microphase separation in block copolymers},}\ }\href@noop {} {\bibfield
  {journal} {\bibinfo  {journal} {Macromolecules}\ } (\bibinfo {year}
  {1980})}\BibitemShut {NoStop}%
\bibitem [{\citenamefont {Fortini}\ \emph {et~al.}(2016)\citenamefont
  {Fortini}, \citenamefont {Mart{\'i}n-Fabiani}, \citenamefont {Lesage De
  La~Haye}, \citenamefont {Dugas}, \citenamefont {Lansalot}, \citenamefont
  {D'Agosto}, \citenamefont {Bourgeat-Lami}, \citenamefont {Keddie},\ and\
  \citenamefont {Sear}}]{Fortini_2016PRL}%
  \BibitemOpen
  \bibfield  {author} {\bibinfo {author} {\bibfnamefont {A.}~\bibnamefont
  {Fortini}}, \bibinfo {author} {\bibfnamefont {I.}~\bibnamefont
  {Mart{\'i}n-Fabiani}}, \bibinfo {author} {\bibfnamefont {J.}~\bibnamefont
  {Lesage De La~Haye}}, \bibinfo {author} {\bibfnamefont {P.-Y.}\ \bibnamefont
  {Dugas}}, \bibinfo {author} {\bibfnamefont {M.}~\bibnamefont {Lansalot}},
  \bibinfo {author} {\bibfnamefont {F.}~\bibnamefont {D'Agosto}}, \bibinfo
  {author} {\bibfnamefont {E.}~\bibnamefont {Bourgeat-Lami}}, \bibinfo {author}
  {\bibfnamefont {J.~L.}\ \bibnamefont {Keddie}},\ and\ \bibinfo {author}
  {\bibfnamefont {R.~P.}\ \bibnamefont {Sear}},\ }\bibfield  {title} {\enquote
  {\bibinfo {title} {Dynamic stratification in drying films of colloidal
  mixtures},}\ }\href {https://doi.org/10.1103/PhysRevLett.116.118301}
  {\bibfield  {journal} {\bibinfo  {journal} {Physical Review Letters}\
  }\textbf {\bibinfo {volume} {116}},\ \bibinfo {pages} {118301} (\bibinfo
  {year} {2016})}\BibitemShut {NoStop}%
\bibitem [{\citenamefont {Zhou}, \citenamefont {Jiang},\ and\ \citenamefont
  {Doi}(2017)}]{Zhou_2017PRL}%
  \BibitemOpen
  \bibfield  {author} {\bibinfo {author} {\bibfnamefont {J.}~\bibnamefont
  {Zhou}}, \bibinfo {author} {\bibfnamefont {Y.}~\bibnamefont {Jiang}},\ and\
  \bibinfo {author} {\bibfnamefont {M.}~\bibnamefont {Doi}},\ }\bibfield
  {title} {\enquote {\bibinfo {title} {Cross interaction drives stratification
  in drying film of binary colloidal mixtures},}\ }\href
  {https://doi.org/10.1103/PhysRevLett.118.108002} {\bibfield  {journal}
  {\bibinfo  {journal} {Physical Review Letters}\ }\textbf {\bibinfo {volume}
  {118}},\ \bibinfo {pages} {108002} (\bibinfo {year} {2017})}\BibitemShut
  {NoStop}%
\bibitem [{\citenamefont {Schulz}\ and\ \citenamefont
  {Keddie}(2018)}]{SchulzKeddie_2018SoftMatter}%
  \BibitemOpen
  \bibfield  {author} {\bibinfo {author} {\bibfnamefont {M.}~\bibnamefont
  {Schulz}}\ and\ \bibinfo {author} {\bibfnamefont {J.~L.}\ \bibnamefont
  {Keddie}},\ }\bibfield  {title} {\enquote {\bibinfo {title} {A critical and
  quantitative review of the stratification of particles during the drying of
  colloidal films},}\ }\href {https://doi.org/10.1039/C8SM01025K} {\bibfield
  {journal} {\bibinfo  {journal} {Soft Matter}\ }\textbf {\bibinfo {volume}
  {14}},\ \bibinfo {pages} {6181--6197} (\bibinfo {year} {2018})}\BibitemShut
  {NoStop}%
\bibitem [{\citenamefont {Lang}\ \emph {et~al.}(2000)\citenamefont {Lang},
  \citenamefont {Likos}, \citenamefont {Watzlawek},\ and\ \citenamefont
  {Löwen}}]{Lang_Likos_2000}%
  \BibitemOpen
  \bibfield  {author} {\bibinfo {author} {\bibfnamefont {A.}~\bibnamefont
  {Lang}}, \bibinfo {author} {\bibfnamefont {C.~N.}\ \bibnamefont {Likos}},
  \bibinfo {author} {\bibfnamefont {M.}~\bibnamefont {Watzlawek}},\ and\
  \bibinfo {author} {\bibfnamefont {H.}~\bibnamefont {Löwen}},\ }\bibfield
  {title} {\enquote {\bibinfo {title} {Fluid and solid phases of the gaussian
  core model},}\ }\href@noop {} {\bibfield  {journal} {\bibinfo  {journal} {J.
  Phys.: Condens. Matter}\ }\textbf {\bibinfo {volume} {12}},\ \bibinfo {pages}
  {5087--5108} (\bibinfo {year} {2000})}\BibitemShut {NoStop}%
\bibitem [{\citenamefont {Schweizer}\ and\ \citenamefont
  {Curro}(1997)}]{Schweizer_Curro_1997}%
  \BibitemOpen
  \bibfield  {author} {\bibinfo {author} {\bibfnamefont {K.~S.}\ \bibnamefont
  {Schweizer}}\ and\ \bibinfo {author} {\bibfnamefont {J.~G.}\ \bibnamefont
  {Curro}},\ }\bibfield  {title} {\enquote {\bibinfo {title} {Integral equation
  theories of the structure, thermodynamics, and phase transitions of polymer
  fluids},}\ }\href@noop {} {\bibfield  {journal} {\bibinfo  {journal} {Adv.
  Chem. Phys.}\ }\textbf {\bibinfo {volume} {98}},\ \bibinfo {pages} {1--142}
  (\bibinfo {year} {1997})}\BibitemShut {NoStop}%
\bibitem [{\citenamefont {Chandler}\ and\ \citenamefont
  {Andersen}(1972)}]{Chandler_Andersen_1972}%
  \BibitemOpen
  \bibfield  {author} {\bibinfo {author} {\bibfnamefont {D.}~\bibnamefont
  {Chandler}}\ and\ \bibinfo {author} {\bibfnamefont {H.~C.}\ \bibnamefont
  {Andersen}},\ }\bibfield  {title} {\enquote {\bibinfo {title} {Optimized
  cluster expansions for classical fluids. ii. theory of molecular liquids},}\
  }\href@noop {} {\bibfield  {journal} {\bibinfo  {journal} {J. Chem. Phys.}\
  }\textbf {\bibinfo {volume} {57}},\ \bibinfo {pages} {1930--1937} (\bibinfo
  {year} {1972})}\BibitemShut {NoStop}%
\bibitem [{\citenamefont {Ornstein}\ and\ \citenamefont
  {Zernike}(1914)}]{Ornstein_Zernike_1914}%
  \BibitemOpen
  \bibfield  {author} {\bibinfo {author} {\bibfnamefont {L.~S.}\ \bibnamefont
  {Ornstein}}\ and\ \bibinfo {author} {\bibfnamefont {F.}~\bibnamefont
  {Zernike}},\ }\bibfield  {title} {\enquote {\bibinfo {title} {Accidental
  deviations of density and opalescence at the critical point of a single
  substance},}\ }\href@noop {} {\bibfield  {journal} {\bibinfo  {journal}
  {Proc. K. Akad. Wet.}\ }\textbf {\bibinfo {volume} {17}},\ \bibinfo {pages}
  {793--806} (\bibinfo {year} {1914})}\BibitemShut {NoStop}%
\bibitem [{\citenamefont {Percus}\ and\ \citenamefont
  {Yevick}(1958)}]{PercusYevick1958}%
  \BibitemOpen
  \bibfield  {author} {\bibinfo {author} {\bibfnamefont {J.~K.}\ \bibnamefont
  {Percus}}\ and\ \bibinfo {author} {\bibfnamefont {G.~J.}\ \bibnamefont
  {Yevick}},\ }\bibfield  {title} {\enquote {\bibinfo {title} {Analysis of
  classical statistical mechanics by means of collective coordinates},}\ }\href
  {https://doi.org/10.1103/PhysRev.110.1} {\bibfield  {journal} {\bibinfo
  {journal} {Physical Review}\ }\textbf {\bibinfo {volume} {110}},\ \bibinfo
  {pages} {1--13} (\bibinfo {year} {1958})}\BibitemShut {NoStop}%
\bibitem [{\citenamefont {Ichiye}\ and\ \citenamefont
  {Chandler}(1988)}]{Ichiye1988}%
  \BibitemOpen
  \bibfield  {author} {\bibinfo {author} {\bibfnamefont {T.}~\bibnamefont
  {Ichiye}}\ and\ \bibinfo {author} {\bibfnamefont {D.}~\bibnamefont
  {Chandler}},\ }\bibfield  {title} {\enquote {\bibinfo {title} {{Hypernetted
  chain closure reference interaction site method theory of structure and
  thermodynamics for alkanes in water}},}\ }\href@noop {} {\bibfield  {journal}
  {\bibinfo  {journal} {Journal of Physical Chemistry}\ }\textbf {\bibinfo
  {volume} {92}},\ \bibinfo {pages} {5257--5261} (\bibinfo {year}
  {1988})}\BibitemShut {NoStop}%
\bibitem [{\citenamefont {Hansen}\ and\ \citenamefont
  {Verlet}(1969)}]{HansenVerlet1969}%
  \BibitemOpen
  \bibfield  {author} {\bibinfo {author} {\bibfnamefont {J.-P.}\ \bibnamefont
  {Hansen}}\ and\ \bibinfo {author} {\bibfnamefont {L.}~\bibnamefont
  {Verlet}},\ }\bibfield  {title} {\enquote {\bibinfo {title} {Phase
  transitions of the {L}ennard-{J}ones system},}\ }\href
  {https://doi.org/10.1103/PhysRev.184.151} {\bibfield  {journal} {\bibinfo
  {journal} {Physical Review}\ }\textbf {\bibinfo {volume} {184}},\ \bibinfo
  {pages} {151--161} (\bibinfo {year} {1969})}\BibitemShut {NoStop}%
\bibitem [{\citenamefont {Plimpton}(1995)}]{Plimpton_1995}%
  \BibitemOpen
  \bibfield  {author} {\bibinfo {author} {\bibfnamefont {S.}~\bibnamefont
  {Plimpton}},\ }\bibfield  {title} {\enquote {\bibinfo {title} {Fast parallel
  algorithms for short-range molecular dynamics},}\ }\href@noop {} {\bibfield
  {journal} {\bibinfo  {journal} {J. Comput. Phys.}\ }\textbf {\bibinfo
  {volume} {117}},\ \bibinfo {pages} {1--19} (\bibinfo {year}
  {1995})}\BibitemShut {NoStop}%
\bibitem [{\citenamefont {Nos{\'e}}(1984)}]{Nose1984}%
  \BibitemOpen
  \bibfield  {author} {\bibinfo {author} {\bibfnamefont {S.}~\bibnamefont
  {Nos{\'e}}},\ }\bibfield  {title} {\enquote {\bibinfo {title} {A unified
  formulation of the constant temperature molecular dynamics methods},}\
  }\href@noop {} {\bibfield  {journal} {\bibinfo  {journal} {The Journal of
  Chemical Physics}\ }\textbf {\bibinfo {volume} {81}},\ \bibinfo {pages}
  {511--519} (\bibinfo {year} {1984})}\BibitemShut {NoStop}%
\bibitem [{\citenamefont {Hoover}(1985)}]{Hoover1985}%
  \BibitemOpen
  \bibfield  {author} {\bibinfo {author} {\bibfnamefont {W.~G.}\ \bibnamefont
  {Hoover}},\ }\bibfield  {title} {\enquote {\bibinfo {title} {Canonical
  dynamics: Equilibrium phase-space distributions},}\ }\href@noop {} {\bibfield
   {journal} {\bibinfo  {journal} {Physical Review A}\ }\textbf {\bibinfo
  {volume} {31}},\ \bibinfo {pages} {1695--1697} (\bibinfo {year}
  {1985})}\BibitemShut {NoStop}%
\bibitem [{\citenamefont {Erigi}, \citenamefont {Dhumal},\ and\ \citenamefont
  {Tripathy}(2023)}]{Erigi_2023}%
  \BibitemOpen
  \bibfield  {author} {\bibinfo {author} {\bibfnamefont {U.}~\bibnamefont
  {Erigi}}, \bibinfo {author} {\bibfnamefont {U.}~\bibnamefont {Dhumal}},\ and\
  \bibinfo {author} {\bibfnamefont {M.}~\bibnamefont {Tripathy}},\ }\bibfield
  {title} {\enquote {\bibinfo {title} {Phase behavior of mixtures of hard
  colloids and soft coarse-grained macromolecules.}}\ }\href@noop {} {\bibfield
   {journal} {\bibinfo  {journal} {The Journal of chemical physics}\ }
  (\bibinfo {year} {2023})}\BibitemShut {NoStop}%
\bibitem [{\citenamefont {Likos}, \citenamefont {Watzlawek},\ and\
  \citenamefont {L{\"o}wen}(1998)}]{Likos_1998_PRE_PenetrableSpheres}%
  \BibitemOpen
  \bibfield  {author} {\bibinfo {author} {\bibfnamefont {C.~N.}\ \bibnamefont
  {Likos}}, \bibinfo {author} {\bibfnamefont {M.}~\bibnamefont {Watzlawek}},\
  and\ \bibinfo {author} {\bibfnamefont {H.}~\bibnamefont {L{\"o}wen}},\
  }\bibfield  {title} {\enquote {\bibinfo {title} {Freezing and clustering
  transitions for penetrable spheres},}\ }\href@noop {} {\bibfield  {journal}
  {\bibinfo  {journal} {Physical Review E}\ }\textbf {\bibinfo {volume} {58}},\
  \bibinfo {pages} {3135--3144} (\bibinfo {year} {1998})}\BibitemShut {NoStop}%
\bibitem [{\citenamefont {Likos}\ \emph
  {et~al.}(2001{\natexlab{a}})\citenamefont {Likos}, \citenamefont {Lang},
  \citenamefont {Watzlawek},\ and\ \citenamefont
  {L{\"o}wen}}]{Likos_2001_PRE_Criterion}%
  \BibitemOpen
  \bibfield  {author} {\bibinfo {author} {\bibfnamefont {C.~N.}\ \bibnamefont
  {Likos}}, \bibinfo {author} {\bibfnamefont {A.}~\bibnamefont {Lang}},
  \bibinfo {author} {\bibfnamefont {M.}~\bibnamefont {Watzlawek}},\ and\
  \bibinfo {author} {\bibfnamefont {H.}~\bibnamefont {L{\"o}wen}},\ }\bibfield
  {title} {\enquote {\bibinfo {title} {Exact criterion for determining
  clustering versus reentrant melting behavior for bounded interaction
  potentials},}\ }\href@noop {} {\bibfield  {journal} {\bibinfo  {journal}
  {Physical Review E}\ }\textbf {\bibinfo {volume} {63}},\ \bibinfo {pages}
  {031206} (\bibinfo {year} {2001}{\natexlab{a}})}\BibitemShut {NoStop}%
\bibitem [{\citenamefont {Zhang}, \citenamefont {Charbonneau},\ and\
  \citenamefont {Mladek}(2010)}]{Zhang_Charbonneau_Mladek_2010_PRL}%
  \BibitemOpen
  \bibfield  {author} {\bibinfo {author} {\bibfnamefont {K.}~\bibnamefont
  {Zhang}}, \bibinfo {author} {\bibfnamefont {P.}~\bibnamefont {Charbonneau}},\
  and\ \bibinfo {author} {\bibfnamefont {B.~M.}\ \bibnamefont {Mladek}},\
  }\bibfield  {title} {\enquote {\bibinfo {title} {Reentrant and isostructural
  transitions in a cluster-crystal former},}\ }\href@noop {} {\bibfield
  {journal} {\bibinfo  {journal} {Physical Review Letters}\ }\textbf {\bibinfo
  {volume} {105}},\ \bibinfo {pages} {245701} (\bibinfo {year}
  {2010})}\BibitemShut {NoStop}%
\bibitem [{\citenamefont {Liu}\ and\ \citenamefont
  {Xi}(2019)}]{Liu_Xi_2019_COCIS}%
  \BibitemOpen
  \bibfield  {author} {\bibinfo {author} {\bibfnamefont {Y.}~\bibnamefont
  {Liu}}\ and\ \bibinfo {author} {\bibfnamefont {Y.}~\bibnamefont {Xi}},\
  }\bibfield  {title} {\enquote {\bibinfo {title} {Colloidal systems with a
  short-range attraction and long-range repulsion: Phase diagrams, structures,
  and dynamics},}\ }\href@noop {} {\bibfield  {journal} {\bibinfo  {journal}
  {Current Opinion in Colloid \& Interface Science}\ }\textbf {\bibinfo
  {volume} {39}},\ \bibinfo {pages} {123--136} (\bibinfo {year}
  {2019})}\BibitemShut {NoStop}%
\bibitem [{\citenamefont {Muna{\`o}}\ \emph {et~al.}(2022)\citenamefont
  {Muna{\`o}}, \citenamefont {Costa}, \citenamefont {Malescio}, \citenamefont
  {Bomont},\ and\ \citenamefont {Prestipino}}]{Munao_2022_SoftMatter}%
  \BibitemOpen
  \bibfield  {author} {\bibinfo {author} {\bibfnamefont {G.}~\bibnamefont
  {Muna{\`o}}}, \bibinfo {author} {\bibfnamefont {D.}~\bibnamefont {Costa}},
  \bibinfo {author} {\bibfnamefont {G.}~\bibnamefont {Malescio}}, \bibinfo
  {author} {\bibfnamefont {J.-M.}\ \bibnamefont {Bomont}},\ and\ \bibinfo
  {author} {\bibfnamefont {S.}~\bibnamefont {Prestipino}},\ }\bibfield  {title}
  {\enquote {\bibinfo {title} {Competition between clustering and phase
  separation in binary mixtures containing {SALR} particles},}\ }\href@noop {}
  {\bibfield  {journal} {\bibinfo  {journal} {Soft Matter}\ }\textbf {\bibinfo
  {volume} {18}},\ \bibinfo {pages} {6453--6464} (\bibinfo {year}
  {2022})}\BibitemShut {NoStop}%
\bibitem [{\citenamefont {Mladek}\ \emph {et~al.}(2006)\citenamefont {Mladek},
  \citenamefont {Gottwald}, \citenamefont {Kahl}, \citenamefont {Neumann},\
  and\ \citenamefont {Likos}}]{Mladek_2006_PRL_ClusterPhases}%
  \BibitemOpen
  \bibfield  {author} {\bibinfo {author} {\bibfnamefont {B.~M.}\ \bibnamefont
  {Mladek}}, \bibinfo {author} {\bibfnamefont {D.}~\bibnamefont {Gottwald}},
  \bibinfo {author} {\bibfnamefont {G.}~\bibnamefont {Kahl}}, \bibinfo {author}
  {\bibfnamefont {M.}~\bibnamefont {Neumann}},\ and\ \bibinfo {author}
  {\bibfnamefont {C.~N.}\ \bibnamefont {Likos}},\ }\bibfield  {title} {\enquote
  {\bibinfo {title} {Formation of polymorphic cluster phases for a class of
  models of purely repulsive soft spheres},}\ }\href@noop {} {\bibfield
  {journal} {\bibinfo  {journal} {Physical Review Letters}\ }\textbf {\bibinfo
  {volume} {96}},\ \bibinfo {pages} {045701} (\bibinfo {year}
  {2006})}\BibitemShut {NoStop}%
\bibitem [{\citenamefont {Archer}, \citenamefont {Evans},\ and\ \citenamefont
  {Roth}(2004)}]{Archer2004a}%
  \BibitemOpen
  \bibfield  {author} {\bibinfo {author} {\bibfnamefont {A.~J.}\ \bibnamefont
  {Archer}}, \bibinfo {author} {\bibfnamefont {R.}~\bibnamefont {Evans}},\ and\
  \bibinfo {author} {\bibfnamefont {R.}~\bibnamefont {Roth}},\ }\bibfield
  {title} {\enquote {\bibinfo {title} {The structure of binary mixtures of
  gaussian core particles},}\ }\href@noop {} {\bibfield  {journal} {\bibinfo
  {journal} {Journal of Physics: Condensed Matter}\ }\textbf {\bibinfo {volume}
  {16}},\ \bibinfo {pages} {1135--1148} (\bibinfo {year} {2004})}\BibitemShut
  {NoStop}%
\bibitem [{\citenamefont {Archer}, \citenamefont {Likos},\ and\ \citenamefont
  {Evans}(2004)}]{Archer2004lamda}%
  \BibitemOpen
  \bibfield  {author} {\bibinfo {author} {\bibfnamefont {A.~J.}\ \bibnamefont
  {Archer}}, \bibinfo {author} {\bibfnamefont {C.~N.}\ \bibnamefont {Likos}},\
  and\ \bibinfo {author} {\bibfnamefont {R.}~\bibnamefont {Evans}},\ }\bibfield
   {title} {\enquote {\bibinfo {title} {{Soft-core binary fluid exhibiting a
  $\lambda$-line and freezing to a highly delocalized crystal}},}\ }\href@noop
  {} {\bibfield  {journal} {\bibinfo  {journal} {Journal of Physics Condensed
  Matter}\ }\textbf {\bibinfo {volume} {16}} (\bibinfo {year}
  {2004})}\BibitemShut {NoStop}%
\bibitem [{\citenamefont {Archer}\ and\ \citenamefont
  {Likos}(2007)}]{Archer2007}%
  \BibitemOpen
  \bibfield  {author} {\bibinfo {author} {\bibfnamefont {A.~J.}\ \bibnamefont
  {Archer}}\ and\ \bibinfo {author} {\bibfnamefont {C.~N.}\ \bibnamefont
  {Likos}},\ }\bibfield  {title} {\enquote {\bibinfo {title} {Liquid-liquid
  demixing and solidification in binary mixtures of ultrasoft particles},}\
  }\href@noop {} {\bibfield  {journal} {\bibinfo  {journal} {Journal of
  Physics: Condensed Matter}\ }\textbf {\bibinfo {volume} {19}},\ \bibinfo
  {pages} {375106} (\bibinfo {year} {2007})}\BibitemShut {NoStop}%
\bibitem [{\citenamefont {Wang}\ \emph
  {et~al.}(2025{\natexlab{a}})\citenamefont {Wang}, \citenamefont {Dhumal},
  \citenamefont {Zakhari},\ and\ \citenamefont {Zia}}]{wang2025elusive}%
  \BibitemOpen
  \bibfield  {author} {\bibinfo {author} {\bibfnamefont {J.~G.}\ \bibnamefont
  {Wang}}, \bibinfo {author} {\bibfnamefont {U.}~\bibnamefont {Dhumal}},
  \bibinfo {author} {\bibfnamefont {M.~E.~A.}\ \bibnamefont {Zakhari}},\ and\
  \bibinfo {author} {\bibfnamefont {R.~N.}\ \bibnamefont {Zia}},\ }\href@noop
  {} {\enquote {\bibinfo {title} {The elusive fluid-and-crystal coexistence
  state in simulations of monodisperse, hard-sphere colloids},}\ } (\bibinfo
  {year} {2025}{\natexlab{a}}),\ \Eprint {https://arxiv.org/abs/2412.05422}
  {arXiv:2412.05422 [cond-mat.soft]} \BibitemShut {NoStop}%
\bibitem [{\citenamefont {Wang}\ \emph
  {et~al.}(2025{\natexlab{b}})\citenamefont {Wang}, \citenamefont {Dhumal},
  \citenamefont {Zakhari},\ and\ \citenamefont {Zia}}]{wang2025frenkel}%
  \BibitemOpen
  \bibfield  {author} {\bibinfo {author} {\bibfnamefont {J.~G.}\ \bibnamefont
  {Wang}}, \bibinfo {author} {\bibfnamefont {U.}~\bibnamefont {Dhumal}},
  \bibinfo {author} {\bibfnamefont {M.~E.~A.}\ \bibnamefont {Zakhari}},\ and\
  \bibinfo {author} {\bibfnamefont {R.~N.}\ \bibnamefont {Zia}},\ }\href@noop
  {} {\enquote {\bibinfo {title} {Frenkel's entropy-exchange mechanism in
  monodisperse, nearly hard-sphere colloids: minimal perturbations to access
  fluid-crystal coexistence},}\ } (\bibinfo {year} {2025}{\natexlab{b}}),\
  \Eprint {https://arxiv.org/abs/2510.06506} {arXiv:2510.06506 [cond-mat.soft]}
  \BibitemShut {NoStop}%
\bibitem [{\citenamefont {Likos}\ \emph
  {et~al.}(2001{\natexlab{b}})\citenamefont {Likos}, \citenamefont {Lang},
  \citenamefont {Watzlawek},\ and\ \citenamefont {L{\"{o}}wen}}]{Likos2001a}%
  \BibitemOpen
  \bibfield  {author} {\bibinfo {author} {\bibfnamefont {C.~N.}\ \bibnamefont
  {Likos}}, \bibinfo {author} {\bibfnamefont {a.}~\bibnamefont {Lang}},
  \bibinfo {author} {\bibfnamefont {M.}~\bibnamefont {Watzlawek}},\ and\
  \bibinfo {author} {\bibfnamefont {H.}~\bibnamefont {L{\"{o}}wen}},\
  }\bibfield  {title} {\enquote {\bibinfo {title} {{Criterion for determining
  clustering versus reentrant melting behavior for bounded interaction
  potentials.}}}\ }\href@noop {} {\bibfield  {journal} {\bibinfo  {journal}
  {Physical review. E, Statistical, nonlinear, and soft matter physics}\
  }\textbf {\bibinfo {volume} {63}},\ \bibinfo {pages} {031206} (\bibinfo
  {year} {2001}{\natexlab{b}})}\BibitemShut {NoStop}%
\bibitem [{\citenamefont {Coslovich}, \citenamefont {Bernabei},\ and\
  \citenamefont {Moreno}(2013)}]{Coslovich2013}%
  \BibitemOpen
  \bibfield  {author} {\bibinfo {author} {\bibfnamefont {D.}~\bibnamefont
  {Coslovich}}, \bibinfo {author} {\bibfnamefont {M.}~\bibnamefont
  {Bernabei}},\ and\ \bibinfo {author} {\bibfnamefont {A.~J.}\ \bibnamefont
  {Moreno}},\ }\bibfield  {title} {\enquote {\bibinfo {title} {Clustering and
  reentrant melting in a system of ultrasoft particles},}\ }\href@noop {}
  {\bibfield  {journal} {\bibinfo  {journal} {The Journal of Chemical Physics}\
  }\textbf {\bibinfo {volume} {138}},\ \bibinfo {pages} {12A539} (\bibinfo
  {year} {2013})}\BibitemShut {NoStop}%
\end{thebibliography}

%

\end{document}